\documentclass[aps,prl,reprint,superscriptaddress]{revtex4-2}

\usepackage{bm,amsmath,amssymb}
\usepackage{verbatim}
\usepackage{graphicx}
\usepackage{float}
\usepackage{multirow}
\usepackage{color}
\usepackage{braket}
\usepackage{hyperref}
\usepackage[dvipsnames]{xcolor}
\hypersetup{colorlinks = true, linkcolor={black}, citecolor={black}, urlcolor = blue}
\usepackage{lipsum, babel}

\newcommand{\valuePhi}{0.0133(8)\,\mathrm{s}^{-1}\mathrm{cm}^{-2}}
\newcommand{\valuePhiNoUnits}{0.0133(8)}
\newcommand{\valueCRfraction}{\mbox{$17.1\pm 1.3\%$}}
\newcommand{\valueCoverageNoPercent}{\mbox{$13.3\pm 0.4$}}
\newcommand{\valueCoverage}{\mbox{$13.3\pm 0.4\%$}}
\newcommand{\valueCoverageNoUnc}{13.3\%}
\newcommand{\valueCRrate}{1/(592\substack{+48\\-41}\,\mathrm{s})}
\newcommand{\valueCRrateSeconds}{592\substack{+48\\-41}\,\mathrm{s}}
\newcommand{\valueCRrateExpect}{1/(574\substack{+38\\-33}\,\mathrm{s})}
\newcommand{\valueCRrateNoUnits}{1/(592\substack{+48\\-41})}
\newcommand{\valueCRrateExpectNoUnits}{1/(574\substack{+38\\-33})}
\newcommand{\valueQotherRateNoUnits}{1/(120 \pm 2)}
\newcommand{\valueRateQ}{1/(101\pm1\,\mathrm{s})}
\newcommand{\valueBkgndQD}{1/(41.9\pm0.9\,\mathrm{h})}
\newcommand{\valueRateQD}{1/(72\pm5\,\mathrm{min})}
\newcommand{\valueRateCRQD}{1/(74\substack{+6\\-5}\,\mathrm{min})}
\newcommand{\valueRateQnoUnits}{1/(101\pm1)}
\newcommand{\valueBkgndQDnoUnits}{1/(41.9\pm0.9)}
\newcommand{\valueRateQDnoUnits}{1/(72\pm5)}
\newcommand{\valueRateCRQDnoUnits}{1/(74\substack{+6\\-5})}
\newcommand{\valueNbkgndQD}{6.36\pm0.07}
\newcommand{\valueNcoinQD}{222}
\newcommand{\valueNS}{\mbox{14,403,488}} % 14403697
\newcommand{\valueNQ}{\mbox{9,460}}
\newcommand{\valueNQD}{222}
\newcommand{\valueNentries}{\mbox{62,820}}
\newcommand{\valueEffQD}{94\%} % efficiency of QS detection, from matched filter
\newcommand{\valueEffS}{98.7\,\%} % collective efficiency of detector category S
\newcommand{\valueEff}{92\%} % product of EffS * Effcoin
\newcommand{\valueEffQDnoPercent}{94}
\newcommand{\valueEffSnoPercent}{98.7}
\newcommand{\valueEffnoPercent}{92}
\newcommand{\valueSigmaOtherFractionQ}{14\%}

\newcommand{\valueCycleDurationAvg}{15.274} % weighted avg. cycle duration, also for entry duration
\newcommand{\valueCycleDurationAvgApprox}{15.3} % Approx. weighted avg. cycle duration, also for entry duration
\newcommand{\valueDurationHours}{266.5}
\newcommand{\valueDurationHoursPrecise}{266.531}

\newcommand{\valueSigmaQ}{0.131\,\mathrm{cm}^2}
\newcommand{\valueSigmaQD}{0.0189\,\mathrm{cm}^2}
\newcommand{\valueSigmaQnoUnits}{0.131}
\newcommand{\valueSigmaQDnoUnits}{0.0189}
\newcommand{\valueSigmaQDEpsilonnoUnits}{0.0187} % effective cross-section accounting for detector efficiencies
\newcommand{\Eadc}{{V_\mathrm{ADC}}}
\newcommand{\EMeV}{{E_\mathrm{s}}}
\newcommand{\VtoE}{{a}}
\newcommand{\Eres}{{b}}
\newcommand{\Cs}{{$^\text{137}$Cs}}
\newcommand{\JJplacement}[1]{\mbox{JJ-placement-\uppercase{#1}}}
\newcommand{\JJplacementLetterOnly}[1]{{\uppercase{#1}}}
\newcommand{\stc}{{spatiotemporally correlated}}
\newcommand{\dtcoin}{{\delta t}_\mathrm{QS}}
\newcommand{\dtcycle}{{{\Delta t}_\mathrm{cycle}}}
\newcommand{\dtdelay}{{\Delta t_\mathrm{wait}}}
\newcommand{\adc}{{\sc adc}}
\newcommand{\geant}{{\sc geant4}}

\newcommand{\pois}[2]{\mathrm{Pois}(#1)_{#2}}
\newcommand{\sigmaExA}{\sigma_\mathrm{A}^{*}}
\newcommand{\sigmaExB}{\sigma_\mathrm{B}^{*}}
\newcommand{\ds}{{d\text{-}\mathrm{S}}}
\newcommand{\CQS}{{C_\mathrm{QS}}}
\newcommand{\ND}{{N_\mathrm{S}}}
\newcommand{\NQ}{{N_\mathrm{Q}}}
\newcommand{\NQD}{{N_\mathrm{QS}}}
\newcommand{\rateD}{{r_\mathrm{S}}}
\newcommand{\rateQ}{{r_\mathrm{Q}}}
\newcommand{\rateQD}{{r_\mathrm{QS}}}
\newcommand{\rateCRQ}{{r_\mathrm{Q}^\mu}}
\newcommand{\rateCRQD}{{r_\mathrm{QS}^\mu}}
\newcommand{\bkgndQD}{{r_\mathrm{QS}^\text{\tiny acc.}}}
\newcommand{\NbkgndQD}{{N_\mathrm{QS}^\text{\tiny acc.}}}
\newcommand{\rateOtherQ}{{r_\mathrm{Q}^\text{\tiny other}}}
\newcommand{\sigmaCRQ}{{\sigma_\mathrm{Q}}}
\newcommand{\sigmaCRQD}{{\sigma_\mathrm{QS}}}
\newcommand{\epQD}{{\epsilon}}
\newcommand{\epCoin}{{\epsilon_{\dtcoin}}}
\newcommand{\epD}{{\epsilon_\mathrm{S}}}
\newcommand{\dGammaInit}{{\Delta\Gamma_\mathrm{init}}}

\newcommand\padzero[1]{\ifnum #1 < 10 0\fi #1} % command for text "Run-00"
\newcommand{\run}[1]{{\sc Run}-\padzero{#1}}
\newcommand{\film}[1]{L{#1}}
\newcommand{\qubit}[1]{\mathrm{Q}{#1}}
\renewcommand{\det}[1]{\text{\sc {#1}}}

\def\RLEaffil{Research Laboratory of Electronics, Massachusetts Institute of Technology, Cambridge, MA 02139, USA}
\def\LLaffil{MIT Lincoln Laboratory, Lexington, MA 02421, USA}
\def\Physaffil{Department of Physics, Massachusetts Institute of Technology, Cambridge, MA 02139, USA}
\def\EECSaffil{Department of Electrical Engineering and Computer Science, Massachusetts Institute of Technology, Cambridge, MA 02139, USA}

\begin{document}

%\preprint{}

\twocolumngrid

\title{Synchronous Detection of Cosmic Rays and Correlated Errors \mbox{in Superconducting Qubit Arrays}}


%\author{\mbox{Patrick M. Harrington}}	\address{\RLEaffil}
\author{\mbox{Patrick M. Harrington}}
\email[]{patrickmharrington@gmail.com}
\address{\RLEaffil}
\author{\nolinebreak\mbox{Mingyu Li}}	\address{\Physaffil}
\author{\nolinebreak\mbox{Max Hays}}	\address{\RLEaffil}
\author{\nolinebreak\mbox{Wouter Van De Pontseele}}	\address{\Physaffil}
\author{\nolinebreak\mbox{Daniel Mayer}}	\address{\Physaffil}
\author{\nolinebreak\mbox{H. Douglas Pinckney}}	\address{\Physaffil}
\author{\mbox{Felipe Contipelli}}	\address{\LLaffil}
\author{\mbox{Michael Gingras}}	\address{\LLaffil}
\author{\mbox{Bethany M. Niedzielski}}	\address{\LLaffil}
\author{\mbox{Hannah Stickler}}	\address{\LLaffil}
\author{\mbox{Jonilyn L. Yoder}}	\address{\LLaffil}
\author{\mbox{Mollie E. Schwartz}}	\address{\LLaffil}
\author{\mbox{Jeffrey A. Grover}}	\address{\RLEaffil}
\author{\mbox{Kyle Serniak}}\address{\RLEaffil}\address{\LLaffil}
\author{\mbox{William D. Oliver}}\email[]{william.oliver@mit.edu}\address{\RLEaffil}\address{\Physaffil}\address{\EECSaffil}
\author{\mbox{Joseph A. Formaggio}}\email[]{josephf@mit.edu}\address{\Physaffil}

\begin{abstract}
Quantum information processing at scale will require sufficiently stable and long-lived qubits, likely enabled by error-correction codes~\cite{shor95}.
Several recent superconducting-qubit \mbox{experiments~\cite{wile21, mcew22, thor23}}, however, reported observing intermittent spatiotemporally correlated errors that would be problematic for conventional codes, with ionizing radiation being a likely cause.
Here, we directly measured the cosmic-ray contribution to spatiotemporally correlated qubit errors.
We accomplished this by synchronously monitoring cosmic-ray detectors and qubit energy-relaxation dynamics of \mbox{10 transmon} qubits distributed across a \mbox{${5\!\times\!5\!\times\!0.35\,\,\mathrm{mm}^3}$} silicon chip.
Cosmic rays caused correlated errors at a rate of \mbox{$1/(592\substack{+48\\-41}\,\mathrm{s})$}, accounting for \mbox{$17.1\pm 1.3\%$} of all such events.
Our qubits responded to essentially all of the cosmic rays and their secondary particles incident on the chip, consistent with the independently measured arrival flux.
Moreover, we observed that the landscape of the superconducting gap in proximity to the Josephson junctions dramatically impacts the qubit response to cosmic rays.
Given the practical difficulties associated with shielding cosmic rays~\cite{card23}, our results indicate the importance of radiation hardening---for example, superconducting gap engineering---to the realization of robust quantum error correction.
\end{abstract}

\date{\today}
\maketitle
Ionizing radiation from cosmogenic and terrestrial sources is ever-present in the laboratory environment.
The former includes cosmic rays and their secondary particles (muons, neutrons, etc.), which shower the earth with a continuous flux of high-energy ionizing radiation~\cite{Workman2022b}. 
Terrestrial examples include gamma-ray emission from trace quantities of \mbox{potassium-40} and progeny nuclei of the uranium and thorium decay chains~\cite{theo96}, all arising from isotopes found in common laboratory materials, from the concrete in the walls to metal fixtures and printed circuit boards~\cite{card23, loer24}.
While terrestrial sources of radiation can generally be abated by dense shielding (typically lead) and the careful selection of low-radioactivity materials~\cite{loac16, card23, loer24}, cosmogenic radiation penetrates matter with such incredibly high momentum (\mbox{${\gtrsim 1}\,\mathrm{GeV}/c$}) that it is only significantly attenuated by the overburden present in underground facilities.
The difficulty of shielding cosmic rays thus presents a challenge for solid-state quantum processors.

Ionizing radiation affects the electrical response and quantum coherence of superconducting circuits~\cite{wood69, day03, veps20}.
Radiation ionizes atoms within a circuit substrate, creating energetic electron-hole pairs that relax via a cascade involving electron-hole recombination, secondary charge carriers, and phonons. A portion of the energy imparted to the substrate is thereby transported to the superconducting circuit elements~\cite{kapl76}, where it generates non-equilibrium quasiparticles that alter the circuit performance~\cite{cate11,sern18}.
\begin{figure}[H]
\includegraphics{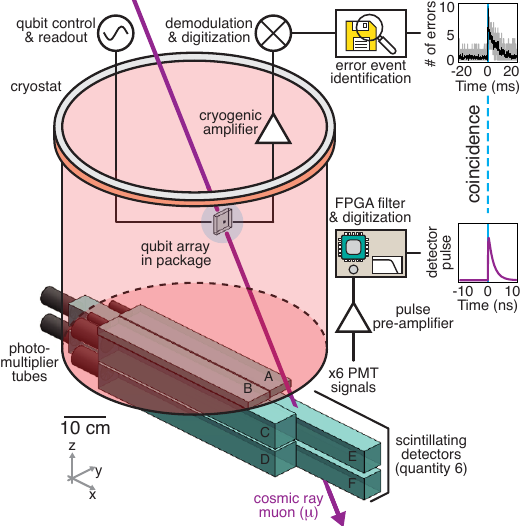}
\caption{\label{fig:fig1p0}
\textbf{Synchronous detection of cosmic rays and qubit relaxation.}
The experiment included a qubit array and scintillating radiation detectors (below cryostat) for continuous monitoring of qubit relaxation and cosmic rays.
We identified individual cosmogenic particles (purple arrow) impacting the qubit array by  coincidence-timing of \stc{} qubit relaxation and detector pulses.
}
\end{figure}

Several groups worldwide have recently studied the impact of ionizing radiation and quasiparticles on the performance of superconducting circuits in the context of quantum information processing.
For example, in the presence of manufactured radioactive sources, excess levels of ionizing radiation increased the average energy-decay rate of superconducting transmon qubits~\cite{veps20}, increased the number of quasiparticle bursts in granular aluminum resonators~\cite{card21}, and increased the rate of quasiparticle-induced phase slips in fluxonium qubits~\cite{guse22}.
The use of lead shielding was shown to reduce qubit decay rates~\cite{veps20} and deep underground operation reduced the rate of both quasiparticle bursts~\cite{card21} and phase slips~\cite{guse22}, which were attributed to the reduction of terrestrial and cosmogenic radiation. Spatiotemporally correlated anomalies in multi-qubit arrays, such as correlated charge-offsets, were also attributed to ionizing radiation~\cite{wile21, iaia22, thor23}.
Additionally, the rate and energy of chip-scale failure events~\cite{mcew22} that inhibited the faithful decoding of quantum error detection protocols~\cite{chen21, acha23} motivates the hypothesis that radiation induces correlated errors, with critical implications for future large-scale applications.

While this body of work is consistent with ionizing radiation being a cause of \stc{} errors, to our knowledge, it has not been shown explicitly that environmental ionizing radiation---whether of terrestrial or cosmogenic origin---is a source of chip-scale qubit errors.
Moreover, the relative contributions of terrestrial and cosmogenic radiation to superconducting qubit errors have been inferred through simulation~\cite{veps20, wile21}, but not yet directly measured. 

In this work, we explicitly identified and quantified \stc{} qubit relaxation events caused by cosmic rays and their secondary particles.
We positioned several scintillating radiation detectors in proximity to a chip with 10 transmon qubits, such that a portion of the cosmogenic radiation flux traverses both \mbox{(Fig.~\ref{fig:fig1p0})}.
We then synchronously monitored multi-qubit relaxation and cosmic-ray detection over a period of \valueDurationHours{} hours, searching for coincident events.
We determined the rate \mbox{$\valueCRrate{}$} of \stc{} events caused by cosmic rays, and furthermore showed that cosmic rays account for \valueCRfraction{} of all such correlated qubit relaxation events.
Finally, we observe that the qubit recovery time following an event is modified by the spatial orientation of the superconducting gap profile around the Josephson junctions relative to the embedding circuitry.
Our results demonstrate that cosmic rays cause a significant fraction of quasiparticle bursts and indicate that gap engineering may alleviate their impact and obviate the need for deep-underground facilities in order to shield superconducting quantum processors from cosmic rays.

\begin{figure}[H]
\includegraphics{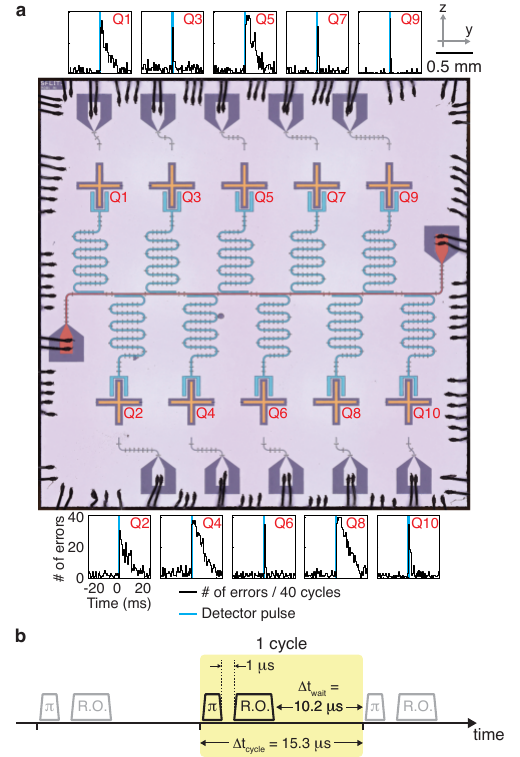}
\caption{\label{fig:fig1p5}
\textbf{Energy-relaxation dynamics from a cosmic ray and the measurement pulse sequence.}
(a) A false-colored micrograph of the qubit array with 10 aluminum transmon qubits (orange), each capacitively coupled to a readout resonator (blue).
Microwave signals for qubit control and readout were routed through a common feedline (red).
The energy-relaxation response of individual qubits to a cosmic-ray impact
(identified in \mbox{Figure~\ref{fig:fig1p0}})
have chip-scale spatial correlations for multiple milliseconds. 
(b) Qubit relaxation events are identified from repeated measurement cycles of a pulse sequence for qubit control and readout.
}
\end{figure}

\section{Experimental Setup}\label{sec:expt_setup}
The qubit device under test (denoted by \det{q}) was an array of 10 fixed-frequency  transmon qubits~\mbox{(Fig.~\ref{fig:fig1p5}a)}.
Each transmon qubit comprises a Josephson junction shunted by a capacitor to a ground plane patterned on a silicon substrate (\mbox{$5\!\times\!5\!\times\!0.35\,\mathrm{mm}^3$}).
Occupation of each qubit's ground and excited states was determined by single-shot interrogation of the qubit-state-dependent response of separate readout resonators~\cite{blai21}.

Identification of \stc{} error events required sampling transient ({\mbox{$\sim 1\,\mathrm{ms}$}}) changes of each qubit's energy-relaxation rate $\Gamma_1$ when it \mbox{exceeded $\sim 1/(1\,\mu\mathrm{s})$}~\cite{mcew22}.
A single instance of qubit relaxation was recorded whenever a qubit was found in its ground state after preparation in the excited state.
Such single relaxation errors were measured for each qubit using repeated cycles of a 
pulse sequence for control and state readout~\mbox{(Fig.~\ref{fig:fig1p5}b)}.
During each of the repeated cycles, the sequence included state preparation ($\pi$-pulse), a fixed \mbox{$1\text{-}\mu\mathrm{s}$} delay, readout, and a \mbox{$10.2\text{-}\mu\mathrm{s}$} wait-time before the following cycle.
Instances of relaxation were likely if a qubit's inverse decay-rate change was comparable to the \mbox{$1\text{-}\mu\mathrm{s}$} delay duration (\mbox{$1/\Gamma_1=T_1\lesssim 1\,\mu\mathrm{s}$}).
Relaxation rate fluctuations could be monitored with sufficient sampling within a duration of \mbox{$\sim1\,\mathrm{ms}$}
since the pulse sequence was repeated every {\mbox{$\dtcycle = 15.3\,\mu\mathrm{s}$}}.
As each cycle was performed consecutively, the single-shot readout signal not only indicated qubit decay, but also the preparation state for the following measurement cycle~ \mbox{(Section~\ref{sec:pulse_sequence})}.

After monitoring qubit relaxation for continuous periods of $10^6$ measurement cycles \mbox{($\approx 15\,\mathrm{s}$)}, we searched for \stc{} events.
We identified events by evaluating a cross-correlation time-series between instances of qubit relaxation (summed over all qubits for each cycle) and an expected temporal evolution of qubit relaxation rates during an event~\mbox{(Section~\ref{sec:mf})}.
The expected temporal correlation of total qubit relaxation was defined as a one-sided exponential with a \mbox{$5\,\mathrm{ms}$} recovery time-constant.
The onset of each event was recognized as a transient peak in this cross-correlation above a threshold value.
The example event shown in \mbox{Figure~\ref{fig:fig1p0}} resembles the expected temporal evolution, and the corresponding relaxation rate dynamics of each qubit (Fig.~\ref{fig:fig1p5}a) exhibits a sudden increase and exponential-like recovery.

The qubit array and detectors were measured for \valueDurationHours{} hours total between \mbox{2023-06-07} and \mbox{2023-06-29}, amounting to {62.82} billion measurement cycles.
After data collection, we identified the timestamps of \stc{} qubit relaxation event arrival times \mbox{$\{t_k\}$} for a total of \valueNQ{} events.
Our observed rate of events, {\mbox{$\rateQ=\valueRateQ{}$}}, is similar to other published results~\cite{wile21, mcew22, thor23}, considering the expected contribution of ionizing radiation impacts due to substrate size of qubit arrays and given the general variability of radiation levels among laboratory environments~\cite{card23, loer24}.

We identified signatures of cosmic rays in the qubit data using synchronized measurements of cosmic rays from six scintillating detectors \mbox{(collectively denoted by {\sc s})} positioned under the experiment cryostat \mbox{(Fig.~\ref{fig:fig1p0})}.
Each detector is a rectangular prism of scintillating polymer optically coupled to a photomultiplier tube (PMT)~(Section~\ref{sec:detectors}).
The PMTs produce a \mbox{$10\text{-}\mathrm{ns}$} pulse of current that is commensurate with the energy deposited in the scintillator from ionizing radiation impacts. 
We used a multi-channel analog-to-digital converter to record detector pulse arrival times and amplitudes.
We ensured selective detection of cosmic rays by only accepting pulse amplitudes within a calibrated range~\mbox{(Section~\ref{sec:det_cal_subsec})}.

All six detectors were monitored concurrently during qubit measurements.
Processing of detector data involved the assignment of each pulse arrival to its contemporaneous qubit measurement cycle.
We observed \valueNS{} measurement cycles during which at least one detector pulse occurred.
This corresponds to an occurrence rate of \mbox{$\rateD = 1/(66.616 \pm 0.017\,\mathrm{ms})$}, which is consistent with our modeled rate of cosmic-ray energy depositions in these scintillating detectors.

\section{Cosmic-ray coincidence identification}\label{sec:coin}
We identified the contribution of cosmic rays to qubit relaxation events by analyzing temporal correlations between the detector and qubit datasets.
From the qubit relaxation event arrival times $\{t_k\}$ and all instances of detector pulses, we calculated each inter-arrival delay $\{\Delta t_k\}$ between a given qubit relaxation event paired with the nearest-in-time detector pulse from any of the detectors.
A coincidence occurs if an inter-arrival delay is within a coincidence window \mbox{$|\Delta t_k| \le \dtcoin/2$}. We define the window duration \mbox{$\dtcoin=3\dtcycle$} to maximize the acceptance of coincidences while minimizing accidental coincidences (false-positives)~\mbox{(Section~\ref{sec:qb_det_coin})}.
The example event in \mbox{Figures~\ref{fig:fig1p0}~and~\ref{fig:fig1p5}a}, shows a detector pulse during the same measurement cycle that marks the onset of the qubit event.
Each coincidence of a qubit relaxation event with a detector pulse indicates an event of cosmogenic origin.
\begin{figure*}[!htbp]
\includegraphics{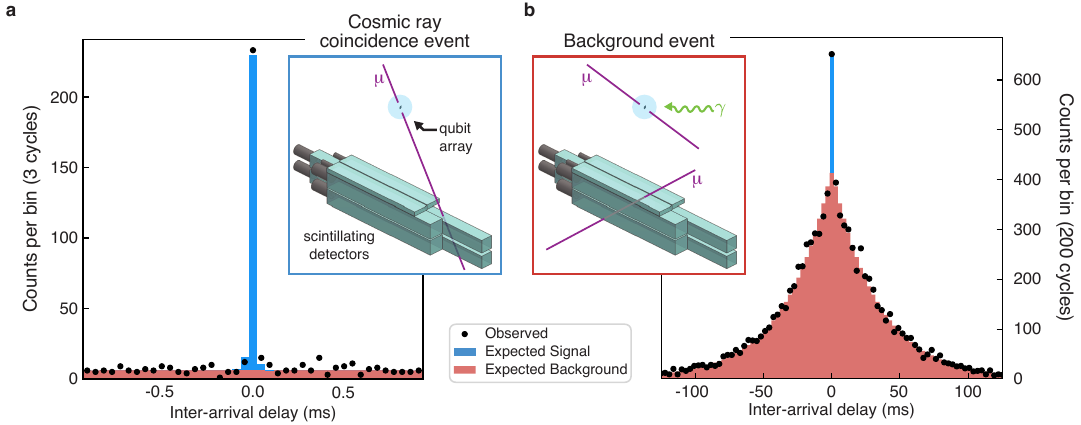}
\caption{\label{fig:fig2}
\textbf{Distribution of inter-arrival delays.}
(a) A histogram of the inter-arrival distribution (\mbox{$3\dtcycle$}-bin-width) shows the observed counts (black), expected background distribution (red), and expected cosmic-ray coincidence counts (blue)~\mbox{(Section~\ref{sec:qb_det_coin})}.
The illustration of a cosmic-ray coincidence event (inset) shows a cosmogenic muon traversing the qubit array and detectors.
(b) The \mbox{$200\dtcycle$}-bin-width histogram of the inter-arrival distribution highlights the agreement between the expected and observed background counts (except in the central bin that also contains cosmic-ray coincidence events).
Example background events (inset) include a muon or gamma ray impacting the qubit array and an unrelated cosmogenic muon traversing the detectors.
}
\end{figure*}

We justify the high likelihood of this claim using the distribution of inter-arrival delays.
\mbox{Figure~\ref{fig:fig2}} displays the inter-arrival distribution using two different histogram bin intervals (\mbox{$3\dtcycle$} and \mbox{$200\dtcycle$}) to depict both the coincidence signal and background components of the inter-arrival distribution.
\mbox{Figure~\ref{fig:fig2}a} shows \valueNcoinQD{} observed qubit-detector coincidences contained in the central bin spanning the coincidence window \mbox{$[-\dtcoin/2, +\dtcoin/2]$}.
The measured coincidence rate {\mbox{$\rateQD=\valueRateQD{}$}} has contributions from both individual cosmic rays and accidental coincidences:
\begin{equation}\label{eq:rateQD}
\underbrace{\rateQD}_{\text{measured}}
=\underbrace{\rateCRQD}_{\text{cosmic-ray}}
+\underbrace{\bkgndQD}_{\text{accidental}}\!\!\!\!,
\end{equation}
where $\rateCRQD$ is the cosmic-ray coincidence rate caused by individual muons (denoted by $\mu$) and $\bkgndQD$ is the rate of accidental coincidences.
Accidental coincidences occur from the random confluence of multiple independent sources \mbox{(Fig.~\ref{fig:fig2}b, inset)}.
We created a background model for the rate of accidental coincidences and all other inter-arrival delays outside the coincidence window.
The expected background distribution was calculated from measured quantities alone (qubit event rate $\rateQ$, the detector pulse rate $\rateD$, and the measured rate of coincidences $\rateQD$) with the consideration that each background inter-arrival delay is not from an individual cosmogenic muon~(Section~\ref{sec:interarrival_bkgnd}).
The background distribution has a characteristic two-sided exponential shape
from the spurious correlation between two independent Poisson processes, which
is visually noticeable when the inter-arrival distribution is binned with \mbox{$200\dtcycle$} intervals \mbox{(Fig.~\ref{fig:fig2}b)}.
We find excellent agreement between the observed inter-arrival distribution and the background model among all inter-arrival delays outside the coincidence window.
The background model prediction within the coincidence window gives an accidental coincidence rate {\mbox{$\bkgndQD=\valueBkgndQD{}$}}, implying that {\mbox{$\valueNbkgndQD{}$}} of the \valueNcoinQD{} coincidence events are accidentals~\mbox{(Fig.~\ref{fig:fig2}a)}. 
We find the rate of coincidences from individual cosmic rays \mbox{(Eq.~\ref{eq:rateQD})} is {\mbox{$\rateCRQD=\valueRateCRQD{}$}}.
Since accidental coincidences are relatively rare (\mbox{$\bkgndQD \ll \rateQD$}), we have high confidence that any given coincidence is from a shared source among the qubit array and detectors, namely a cosmogenic particle. 

\section{The Rate of Qubit Relaxation \mbox{Events from Cosmic Rays}}\label{sec:rate}
Cosmic rays are only one source of \stc{} qubit relaxation events:
\begin{equation}\label{eq:rateQ}
\rateQ = \rateCRQ + \rateOtherQ,
\end{equation}
where $\rateCRQ$ is the rate of events caused by cosmic rays, and $\rateOtherQ$ is the rate of events caused by other sources such as terrestrial radiation.
Note the difference between $\rateCRQ$ and $\rateCRQD$; the former requires an energy deposition to the qubit array, while the latter requires coincident energy depositions to both the qubit array and a detector.

Most cosmic rays incident on the qubit array were not registered as coincidences, because the detectors do not surround or otherwise provide a full coverage of the qubit array.
Nevertheless, the cosmic-ray coincidences $\rateCRQD$ are a known portion $\CQS$ of all
qubit relaxation events from cosmic rays $\rateCRQ$, and we can estimate the rate $\rateCRQ$ from the relationship,
\begin{equation}\label{eq:coverage}
\rateCRQD=\CQS\cdot\rateCRQ,
\end{equation}
where \mbox{$\CQS=\text{\valueCoverage{}}$} is the coverage provided by the detectors.
The coverage was calculated as, \mbox{$\CQS=\epQD\sigmaCRQD/\sigmaCRQ$},
where \mbox{$\sigmaCRQ=\valueSigmaQ{}$} is the cross-section for a cosmic ray to impact the qubit array, \mbox{$\sigmaCRQD=\valueSigmaQD{}$} is the cross-section for a cosmic ray to impact the qubit array and also deposit energy in the detectors within the energy range for pulse acceptance, and \mbox{$\epQD=\valueEff{}$} is the efficiency of qubit-detector coincidence identification~(Section~\ref{sec:qb_det_coin}).

\begin{figure}[!htbp]
\includegraphics{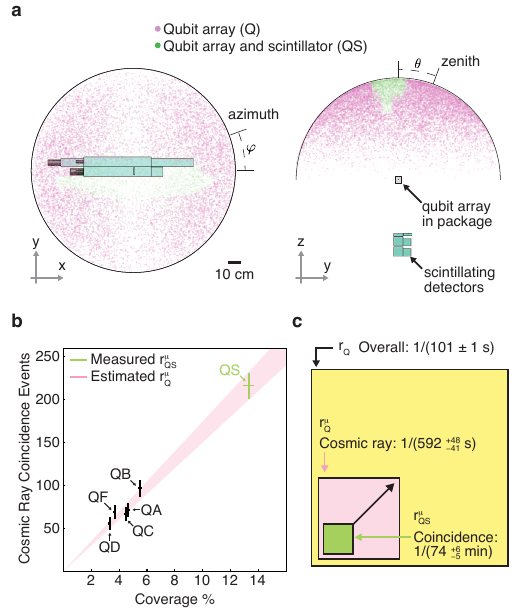}
\caption{\label{fig:fig3}
\textbf{The rate of \stc{} qubit relaxation events from cosmic rays.}
(a) Simulation of cosmic rays in \geant{} informed the cross-section of energy depositions 
to the qubit array substrate and detectors.
The left and right panels show the originating azimuthal and zenith angles for cosmic-ray muons that deposit energy into the qubit array (\det{q}, pink) and the into both the qubit array and any of the detectors (\det{qs}, green).
The relative number of \det{qs} and \det{q} samples informs the coverage---the portion of all cosmic-ray impacts to the qubit array that are also cosmic-ray coincidence events.
(b) The rate of cosmic-ray coincidences depends on the coverage provided by the detectors.
The rate of \stc{} qubit relaxation events from cosmic rays, $\rateCRQ$  (pink), shown as a proportionality factor, was calculated from the observed coincidence counts \det{qs} (green) and the coverage (Eq.~\ref{eq:coverage}).
We overlay selected coincidence combinations (black), which are each consistent with the total coincidence rate.
(c) The rates of cosmic-ray coincidences $\rateCRQD$, cosmic-ray events $\rateCRQ$, and all events $\rateQ$ are represented by overlapping areas.
The overall rate of qubit events (yellow) has contributions from cosmic rays and other sources. The cosmic-ray coincidences (green) are a known portion (\mbox{$\CQS=\valueCoverage{}$}) of all cosmic-ray-induced events (pink), which account for \valueCRfraction{} of the overall event rate.
}
\end{figure}

Cross-sections were based on the numerical sampling of cosmic-ray interactions with the detectors and qubit array using \geant{}~\cite{alli16}.
The calculation of each cross-section accounts for the interdependence among the cosmic-ray flux distribution, geometric arrangement of each detector, and the deposited energy in the detectors.
The hemisphere in \mbox{Figure~\ref{fig:fig3}a} shows the angular position of cosmic-ray muons that were numerically sampled to calculate the interaction cross-sections for impacts to the qubit array (pink, \det{q}) and impacts to both the qubit array and any of the detectors (green, \det{qs}).
The coverage of the qubit array that is collectively provided by detectors is represented by the relative number of points for \det{q} and \det{qs} in the hemisphere \mbox{(Fig.~\ref{fig:fig3}a)}.
We have also calculated cross-sections for coincidence combinations among the six detectors themselves and found they accurately predict their observed energy spectra and coincidence rates~\mbox{(Section~\ref{sec:cr_model}~and~\ref{sec:observation_summary})}.

The rate of \stc{} qubit relaxation events caused by cosmic rays, \mbox{$\rateCRQ=\valueCRrate$}, is calculated from \mbox{Equation~\ref{eq:coverage}}.
This rate is shown as the slope of the confidence band in \mbox{Figure~\ref{fig:fig3}b}, which also plots the number of cosmic-ray coincidences for selected combinations of detectors and the qubit array.
The data point {\sc qs} (green) includes all observed coincidences in the experiment.
We highlight the accuracy and predictive power of the cross-section model by decomposing the observed coincidences into sub-categories (black) among the detectors ({\sc qa}, {\sc qb}, etc.), and find these are each consistent with the overall coincidence rate.
Additionally, there is close agreement between the cosmic-ray event rate $\rateCRQ$ and an expected rate of cosmic-ray impacts \mbox{$\sigmaCRQ\Phi=\valueCRrateExpect$},
based on the measured cosmic-ray flux in the laboratory, \mbox{$\Phi = \valuePhi{}$}~(Section~\ref{sec:eff_flux}).
We conclude that likely all cosmic-ray impacts resulted in a detectable event of \stc{} qubit relaxation for this device.

We found cosmic rays account for \valueCRfraction{} of all the \stc{} events detected.
The remaining events are likely from ionizing radiation impacts from gamma rays sourced in the laboratory and the experiment apparatus.
Phonon and quasiparticle burst events may also result from the absorption of non-ionizing radiation, such as luminescence, Cherenkov radiation, and transition radiation, which is generally induced by ionizing radiation~\cite{adar22, alba22, berg22, peiz22, ponc23}.
Potential non-radiation contributions to the \stc{} event rate may include stress-relaxation~\cite{anth22} and mechanical impulses, e.g. from the pulse-tube cryocooler~\cite{kono23}.

\section{Severity of Spatiotemporal Correlations from Cosmic Rays}
\begin{figure}[!htbp]
\includegraphics{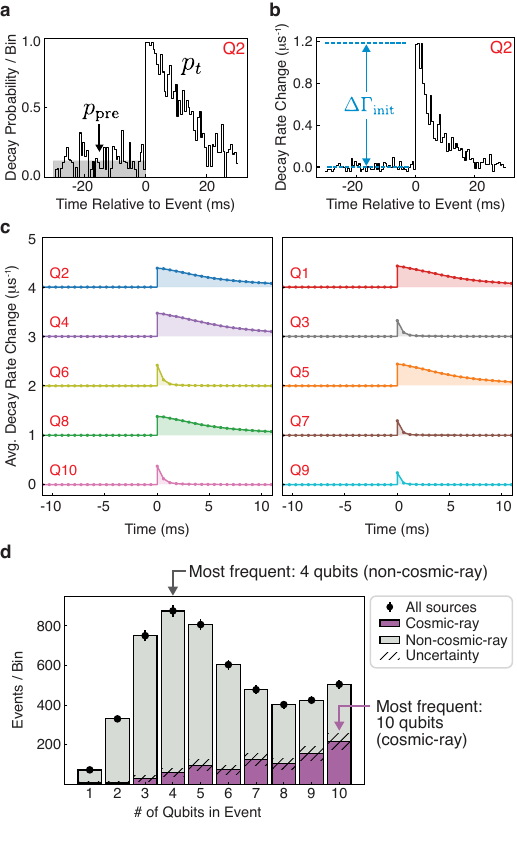}
\caption{\label{fig:fig4}
\textbf{Severity of \stc{} relaxation events.}
(a) The probability of relaxation is calculated for each qubit within each event. As an example, we show the decay probability of Q2 for an example event.
The probabilities $p_\mathrm{t}$ are evaluated for each bin of {40} measurement cycles (black).
The baseline decay probability $p_\mathrm{pre}$ was evaluated from a single pre-trigger bin of {1880} single-shots (gray).
(b) The decay-rate change during this event displays a rapid onset and {\mbox{$\approx 6\,\mathrm{ms}$}} timescale recovery.
The initial decay-rate change, $\Delta\Gamma_\mathrm{init}$ was evaluated to determine the participation of each qubit within each event.
(c) The average decay-rate change over all events shows two distinct timescales of recovery among the qubits.
Traces are incrementally offset by \mbox{$1\,\mu\mathrm{s}^{-1}$}.
(d) The number of qubits participating in each event was based on thresholding the initial decay rate (\mbox{$\dGammaInit \ge 1/(5\,\mu\mathrm{s})$}).
The error bars indicate counting statistics for the total in each bin.
Stacked histograms show the relative contribution from cosmic rays (purple) and other sources (gray), as calculated from coincidence measurements.
Uncertainty in the constructed distributions is based on the counting statistics of coincidence events.
}
\end{figure}

We analyzed the dynamics of qubit relaxation rates during \stc{} events in terms of temporal and spatial correlations.
For each event, we estimated the change in qubit decay rates and characterized their recovery dynamics.
We analyzed each qubit individually by binning the single-shot measurement results in time.
The decay probability within each time bin is
\begin{equation}\label{eq:p_decay}
p = \frac{n_\mathrm{decay}}{n_\mathrm{prep}},
\end{equation}
where $n_\mathrm{prep}$ is the number of preparations and $n_\mathrm{decay}$ is the number of decays within the bin.
The decay probability relates to a decay rate as
\begin{equation}\label{eq:p_exp}
p = 1 - Ae^{-\Gamma\Delta t},
\end{equation}
where $\Gamma$ is the decay rate, \mbox{$\Delta t=3\,\mu\mathrm{s}$} is the effective delay time between qubit state preparation and measurement, and $A$ is a constant related to preparation and measurement fidelity.
We used \mbox{1,880} pre-trigger measurement cycles (\mbox{$\approx 29\,\mathrm{ms}$} prior to the event onset) to evaluate, $p_\mathrm{pre}$, a baseline probability of relaxation (Eq.~\ref{eq:p_decay}).
We also evaluated the decay probability using shorter duration time bins (\mbox{40 cycles $\approx 0.7\,\mathrm{ms}$}) to capture the dynamics of decay-rate fluctuations and recovery.
\mbox{Figure~\ref{fig:fig4}a} displays the decay probability of $\qubit{2}$ during an example event.
We show the pre-trigger baseline probability $p_\mathrm{pre}$ (gray) and the 40-cycle bins, labeled $p_t$, both before and after the event onset.
We calculated (Eq.~\ref{eq:p_exp}) the decay-rate change $\Delta\Gamma_t$ relative to the pre-trigger baseline, as shown in \mbox{Figure~\ref{fig:fig4}b} for for the pre- and post-trigger time bins.

Temporal correlations within an event were summarized in terms of
a time constant $\tau_i$ (of each qubit $i$) for the decay rate recovery to baseline.
Each qubit exhibits a recovery time constant that is consistent from event to event.
The average recovery dynamics for each qubit \mbox{(Fig.~\ref{fig:fig4}c)} clearly have two distinct timescales among the qubits: five qubits have a slow (\mbox{$\tau\approx6\,\mathrm{ms}$}) recovery while the other five qubits have a fast (\mbox{$\tau\approx0.7\,\mathrm{ms}$}) recovery.
The recovery timescales are directly related to the orientation of the Josephson junction electrodes relative to the aluminum ground plane of the qubit array~(Section~\ref{sec:qubit_device}).
The origin of these differences is likely due to the influence of the superconducting gap structure near the Josephson junction on quasiparticle dynamics, though thorough elucidation will be the focus of future work~(Section~\ref{sec:jjorientation}).

We characterized the scale of spatial correlations in terms of the number of qubits participating in each event.
Here, we analyzed the latter {147.1} hours of data for which all 10 qubits were measured~(Section~\ref{sec:summary_runs}).
We defined a qubit to participate in an event if its initial decay-rate change (example indicated in \mbox{Figure~\ref{fig:fig4}b}) exceeded a threshold \mbox{$\dGammaInit \ge 1/(5\,\mu\mathrm{s})$}, which was chosen to limit false-positive assignment.
The likelihood that a given qubit participated in any given event ranges from \mbox{$47\%-67\%$} (and is not directly related to Josephson junction placement).
\mbox{Figure~\ref{fig:fig4}d} shows the distribution of the number of qubits participating in each event (black points) from both cosmic-ray and non-cosmic-ray sources.
The distribution is bimodal, having a tendency for 4-qubit and 10-qubit events to occur.
Events that have more than eight qubits participating may result from a sensitivity limitation due to qubit-count and variability of qubit sensitivity; events that could affect more than 10 qubits are included in the 9-qubit and 10-qubit bins.
This is reasonable if one considers that the spatial extent of an event can be a proxy for the amount of energy deposited in the qubit array substrate from ionizing radiation sources, which is expected to have a distribution with an exponentially-decreasing tail as energy increases. 

We have established high confidence that a given qubit-detector coincidence event was of  cosmogenic origin (Section~\ref{sec:coin}) and effectively all cosmic-ray impacts resulted in a detected \stc{} event (Section~\ref{sec:rate}).
Accordingly, the qubit relaxation dynamics from coincidence events are representative of all cosmic-ray-induced \stc{} events.
\mbox{Figure~\ref{fig:fig4}d} shows stacked histograms for cosmic-ray (purple) and non-cosmic-ray (gray) contributions to the qubit participation in events.
The cosmic-ray distribution of \mbox{Figure~\ref{fig:fig4}d} was constructed by scaling the measured coincidence counts per bin by the inverse coverage (1/\valueCoverageNoUnc{}) of the detectors.
We found that cosmic-ray events have all qubits participate most frequently.
In comparison, four qubits most frequently participate in events from non-cosmic-ray sources.
These results suggest that cosmic rays tend to cause correlated errors of greater spatial extent compared to events from non-cosmic-ray sources.
This is consistent with the expected deposited energy in the qubit array from background gamma rays and cosmic rays: most cosmic rays deposit \mbox{$\gtrsim 80\,\mathrm{keV}$} while energy depositions from most ambient gamma rays are \mbox{$\mathcal{O}(1\,\mathrm{keV})$}~\cite{veps20}.

\section{Conclusion and outlook}
We have shown that cosmic rays cause superconducting qubit errors.
This was achieved by adopting coincidence-timing techniques that correlate cosmic-ray detection events with changes of the energy-decay rates in a 10-transmon qubit array.
The measured rates of cosmic-ray-induced errors are consistent with interaction cross-sections calculated from a \geant{} model and the independently measured muon flux in the laboratory.
To within statistical certainty, all cosmic rays incident on the qubit array resulted in detectable \stc{} error events.
Cosmic rays contribute to a significant fraction (\valueCRfraction{}) of all \stc{} error events that occur, although the majority of such events are of non-cosmogenic origins, such as gamma-ray sources near the qubit array or within the laboratory environment.
Furthermore, cosmic-ray events were most likely to affect all 10 qubits in the array, whereas non-cosmogenic sources peaked at four qubits. 

Low-background radiation environments may be helpful for understanding overall device susceptibility to ionizing radiation and could be one means to achieve robust operation of real-time quantum error correction.
However, while underground facilities may protect quantum devices~\cite{form04, bert23} and advance scientific knowledge, it would be advantageous from a practical standpoint to develop design and fabrication techniques that mitigate the impact of ionizing radiation on solid-state quantum devices operated above ground.
For example, radiation-hardened superconducting qubits may obviate a need for operation in low-background underground facilities.

Radiation-hardened superconducting qubits could be realized by several methods, including phonon trapping~\cite{henr19, kara19, devi21, iaia22}, quasiparticle trapping~\cite{gold90, wang14, mart21}, or inhibiting quasiparticle tunneling~\cite{aume04, yama06, cour08, kala20}.
In our experiment, we varied the spatial structure of the superconducting gap near the Josephson junctions~\cite{cate22} to alter the observed quasiparticle recovery timescale. 
Further engineering of the superconducting gap at the Josephson junction electrodes and beyond may be able to sufficiently suppress the excess quasiparticle tunneling that arises from ionizing radiation events~\cite{conn23}. 

More studies are needed to understand possible tertiary effects of ionizing radiation on superconducting qubits, such as restructuring the density of states of two-level systems, which in turn cause qubit relaxation~\cite{thor23} and may be responsible for relaxation time $T_1$ fluctuations~\cite{klim18, burn19, schl19, degr20}.
Through this effect, radiation impacts are a source of spatially correlated quasi-static noise that could be problematic for stable performance of quantum error correction. 
Radiation hardening may help mitigate such tertiary effects as well.

Furthermore, error correction protocols can be tailored to detect and accommodate \stc{} errors.
Recently proposed schemes adapt error correction codes around specific error-prone qubits~\cite{auge17, stri23} and mid-circuit anomalies~\cite{suzu22, sane23}, or they spatially separate physical-qubit chiplets~\cite{xu22}.
A complementary approach is to embed radiation sensors within qubit arrays~\cite{orre21, hays23} to flag that an error may have occurred.
Each of these protocols benefits from physical hardware platforms that remain resilient in the presence of ionizing radiation. 

\section*{Acknowledgements}
\begin{acknowledgments}
We acknowledge Ben~Loer, Ray~Bunker, Mike~Kelsey, and John~Orrell for discussions; Niv~Drucker, Kevin~A.~Villegas, Nikola~{\v S}ibali\' c, and Tomer~Feld for Quantum~Machines hardware support; Gregory~Calusine, Aranya~Goswami, Or~Hen, Cyrus~F.~Hirjibehedin, Mallika~T.~Randeria, and Lindley~Winslow for helpful conversations; Joseph~Alongi and Amir~H.~Karamlou for preliminary measurements; Katrina~Li for verifying the geometric placement of detectors via a theodolite. 

This research was supported in part by
the Army Research Office under Award \mbox{No. W911NF-23-1-0045},
the U.S. Department of Energy under Award \mbox{No. DE-SC0019295}, and under the Air Force Contract \mbox{No. FA8702-15-D-0001}.
This research was supported by an appointment to the Intelligence Community Postdoctoral Research Fellowship Program at MIT administered by Oak Ridge Institute for Science and Education (ORISE) through an interagency agreement between the U.S. Department of Energy and the Office of the Director of National Intelligence (ODNI).
Any opinions, findings, conclusions or recommendations expressed in this material are those of the authors and do not necessarily reflect the views of
the Army Research Office, the U.S. Department of Energy,
the U.S. Air Force, or the U.S. Government.
\end{acknowledgments}

\section*{Author Contributions}
PMH, WVDP, and DM conceived the original idea of the experiment.
PMH and ML developed and carried out the experiment and analysis with aid from MH, WVDP, DM, and JAF.
ML performed the simulations.
MG, BMN, and HS fabricated the qubit device with coordination from JLY, MES, and KS.
FC contributed to the interpretation of results.
JAG, KS, WDO, and JAF supervised the project.
PMH wrote the manuscript with support from ML, HDP, JAG, KS, WDO, JAF, and contributions from all authors.

\bibliographystyle{harrington_bibstyle_etal}
\bibliography{references}

\clearpage

\appendix

\onecolumngrid

\section{Supplemental Information}

\begin{table}[htbp]
\caption{\label{tab:symbols}
\textbf{Cosmic-ray rate model parameters.}
}
\begin{tabular}{l r@{\hskip 1.5pt}l l@{\hskip 10pt}rl}
\hline
\multicolumn{1}{c}{Symbol} & \multicolumn{2}{c}{Value} & \multicolumn{1}{c}{Description} & \multicolumn{1}{c}{Evaluation} & \multicolumn{1}{c}{Reference} \\ \hline
$\dtcycle$		& $\valueCycleDurationAvg$ & $\mu\mathrm{s}$ & measurement cycle duration 		& measured & Section~\ref{sec:summary_runs} \\
$\dtcoin$		& $45.821$ & $\mu\mathrm{s}$ & coincidence window									& $3\cdot\dtcycle$ & Section~\ref{sec:qb_det_coin} \\
$n_\mathrm{cycles}$		& $62.82$ &	$\cdot10^9$ & total number of cycles						& defined & Section~\ref{sec:expt_setup}~and~\ref{sec:summary_runs} \\
$T$				& $\valueDurationHoursPrecise$ & $\mathrm{h}$ & total duration of data			& $n_\mathrm{cycles}\cdot\dtcycle$ & Section~\ref{sec:summary_runs}\\
\hline
$\sigmaCRQ$		& $\valueSigmaQnoUnits$ & $\mathrm{cm}^2$ & qubit array (\det{q}) cross-section 							& numerical sampling & Section~\ref{sec:cross_section}~and~\ref{sec:det_cal} \\
$\sigmaCRQD$ 	& $\valueSigmaQDnoUnits$ & $\mathrm{cm}^2$ & qubit-detector (\det{qs}) cross-section	& numerical sampling & Section~\ref{sec:cross_section}~and~\ref{sec:det_cal} \\
$\sigma_\mathrm{QS,\epsilon}$ & $\valueSigmaQDEpsilonnoUnits$ & $\mathrm{cm}^2$ & \det{qs} effective cross-section & Eq.~\ref{eq:rateCRQD} & Section~\ref{sec:coverage} \\
$\epCoin$		& $\valueEffQDnoPercent\pm 1$ & \% & \det{qs} identification efficiency 		& simulated and measured & Section~\ref{sec:qb_det_coin} \\
$\epD$			& $\valueEffSnoPercent$ & \% & efficiency of detectors collectively (\det{s}) 					& $\sigma_\mathrm{QS,\epsilon} / \sigma_\mathrm{QS}$ & Section~\ref{sec:coverage} \\
$\epQD$			& $\valueEffnoPercent$ & \% & net efficiency of \det{qs} detection				& $\epD\epCoin$ & Section~\ref{sec:rate}~and~\ref{sec:coverage} \\
$\CQS$			& \valueCoverageNoPercent{} & \% & coverage of qubit array						& $\epCoin\epD\sigmaCRQD/\sigmaCRQ$ & Section~\ref{sec:rate},~\ref{sec:cross_section},~and~\ref{sec:coverage} \\
\hline
$\NQ$			& $\valueNQ$ & & qubit array (\det{q}) event count								& measured			& Section~\ref{sec:expt_setup},~\ref{sec:mf},~and~\ref{sec:observation_summary} \\
$\ND$			& $\valueNS$ & & collective detector (\det{s}) pulse count						& measured			& Section~\ref{sec:expt_setup}~and~\ref{sec:ref_timing}\\
$\NQD$		 	& $\valueNQD$ & & \det{qs} coincidence count						& measured			& Section~\ref{sec:coin}~and~\ref{sec:qb_det_coin} \\
$\rateQ$		& $\valueRateQnoUnits$ & $\mathrm{s}^{-1}$ & qubit array (\det{q}) event rate								& $\NQ/T$				& Section~\ref{sec:rates} \\
$\rateD$		& $1/66.6$ & $\mathrm{ms}^{-1}$ & collective detector (\det{s}) pulse rate								& $\ND/T$				& Section~\ref{sec:rates} \\
$\rateQD$	 	& $\valueRateQDnoUnits$ & $\mathrm{min}^{-1}$ & \det{qs} coincidence rate						& $\NQD/T$				& Section~\ref{sec:rates} \\
$\bkgndQD$		& $\valueBkgndQDnoUnits$ & $\mathrm{h}^{-1}$ & \det{qs} accidental coincidence rate			& $(\rateQ - \rateQD)(e^{\rateD \dtcoin} - 1)$ & Section~\ref{sec:coin}~and~\ref{sec:interarrival_bkgnd} \\
$\rateCRQD$ 	& $\valueRateCRQDnoUnits$ & $\mathrm{min}^{-1}$ & \det{qs} cosmic-ray coincidence rate			& $\rateCRQD - \bkgndQD$ & Section~\ref{sec:coin}~and~\ref{sec:qb_det_coin} \\
$\rateCRQ$ 		& $\valueCRrateNoUnits$ & $\mathrm{s}^{-1}$  & \det{q} rate from cosmogenic sources				& $\rateCRQD / \CQS$ & Section~\ref{sec:rate}~and~\ref{sec:coverage} \\
$\rateOtherQ$	& $\valueQotherRateNoUnits$ & $\mathrm{s}^{-1}$  & \det{q} rate from non-cosmogenic sources			& $\rateQ - \rateCRQ$ & Section~\ref{sec:rate} \\
\hline
$\Phi$			& $\valuePhiNoUnits$ & $\mathrm{s}^{-1}\mathrm{cm}^{-2}$ & muon flux			& fitted detector-only data & Section~\ref{sec:cr_model_muon}~and~\ref{sec:eff_flux} \\ 
$r_\mathrm{Q}^{\mu,\text{\tiny exp.}}$ & $\valueCRrateExpectNoUnits$ & $\mathrm{s}^{-1}\mathrm{cm}^{-2}$ & expected \det{q} rate from cosmic rays & $\sigmaCRQ\Phi$ & Section~\ref{sec:rate}~and~\ref{sec:cross_section} \\ 
\end{tabular}
\end{table}

\subsection{Experiment Setup}
\subsubsection{Qubit measurement setup}
\begin{figure}[htbp]
\includegraphics{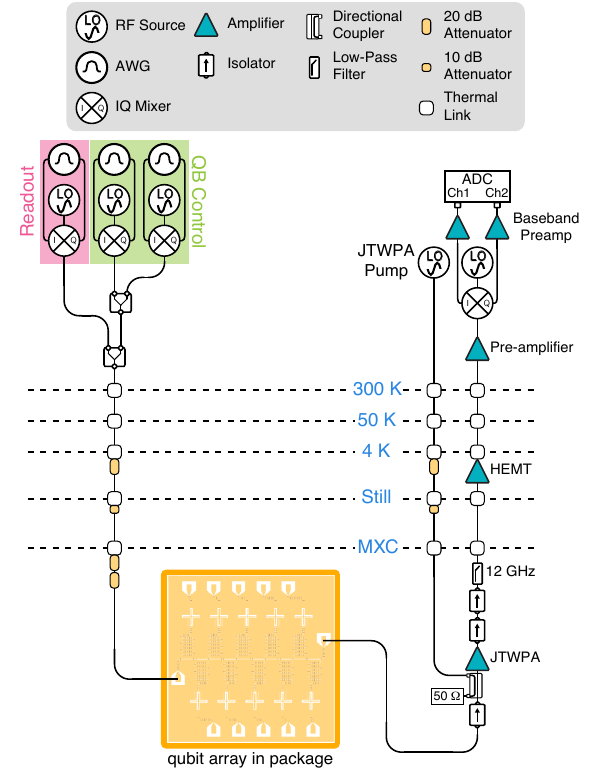}
\caption{\label{fig:fig_wiring} \textbf{Schematic of the qubit measurement setup.} The diagram includes components and connectivity for qubit control and readout. Quantum Machines OPX+ hardware was used for all AWG signals and the analog-to-digital  conversion of qubit single-shot readout signals.
}
\end{figure}

This reported experiment was performed in the same laboratory space (latitude $42.36^\circ\,\mathrm{N}$) and cryostat as the experiments of Veps\"al\"ainen, et al.~\cite{veps20}.
The mixing chamber (MXC) of the Leiden Cryogenics CF-CS81-1500 dilution refrigerator was held at approximately \mbox{$10\,\mathrm{mK}$} throughout data collection.
The qubit array package was mounted to a gold-plated copper paddle attached to the MXC.
For shielding of electromagnetic radiation, the qubit package and the MXC paddle were surrounded by a nested enclosure of superconducting aluminum, tin-plated copper, and high-permeability magnetic shielding (Cryoperm-10).

The wiring setup for qubit measurement is shown in Figure~\ref{fig:fig_wiring}.
The readout signal from the qubit array was first amplified by a Josephson traveling-wave parametric amplifier (JTWPA)~\cite{mack15}, pumped by an Agilent RF source signal joined into the measurement chain with a directional coupler.
The readout signal was further amplified with a high-electron-mobility transistor (HEMT) amplifier at the 4K stage, followed by an amplifier (MITEQ) at room temperature.
After frequency downconversion, the readout signal was further amplified using a Stanford Research SR445A before analog-to-digital conversion using quadrature channels on the Quantum Machines OPX+.

All arbitrary waveform generator (AWG) signals for qubit control and readout pulses were sourced by the Quantum Machines OPX+ hardware and upconverted as single-sideband tones with the internal IQ mixer of the Rohde and Schwarz SGS100A SGMA RF sources.
A reference tone from the ``Readout'' RF source was used as the local oscillator for downconversion of the multiplexed readout signals.
All control electronics were synchronized by a common $10\,\mathrm{MHz}$ rubidium clock source (Stanford Research Systems FS725). 
\begin{table}[htbp]
\caption{\label{tab:equipment}
\textbf{Summary of control equipment.} The manufacturers and model numbers of the qubit measurement-related equipment used for this experiment.
}
\begin{tabular}{l@{\quad}l@{\quad}l}
\hline
\multicolumn{1}{c}{Component} & \multicolumn{1}{c}{Manufacturer} & \multicolumn{1}{c}{Model} \\ \hline
Dilution Refrigerator & Leiden Cryogenics & CF-CS81\\
RF Source (qubit control and readout)& Rohde and Schwarz & SGS100A\\
RF Source (JTWPA pump) & Agilent & E8267D\\
HEMT Amplifier & Low Noise Factory & LNF-LNC0.3\_14A \\
Pre-amplifier & MITEQ & LNA-40-00101200\\
Baseband Pre-amplifier & Stanford Research Systems& SRS445A\\
Baseband Source & Quantum Machines & OPX+\\
Rb Clock & Stanford Research Systems & FS725\\ \hline
\end{tabular}
\end{table}

\subsubsection{Details of the qubit array}\label{sec:qubit_device}
\begin{figure}[htbp]
\includegraphics{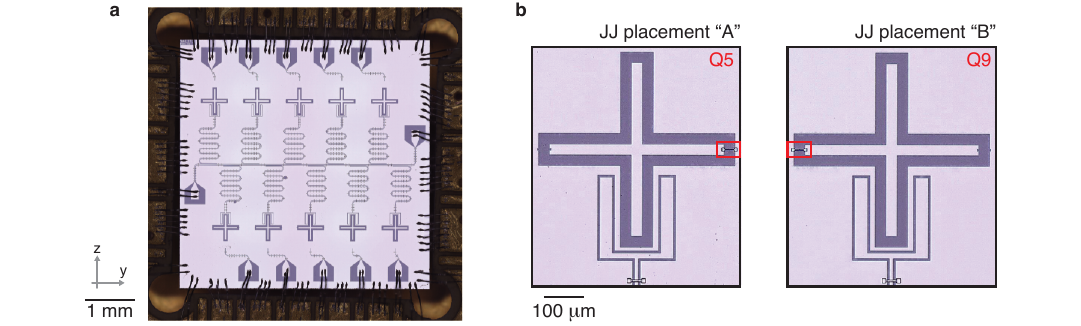}
\caption{\label{fig:fig_device} \textbf{Optical micrographs of the qubit array.} (a) The qubit array mounted in its package by 107 wirebonds and surrounded by the package PCB (TMM10). Labels for each transmon qubit are in Figure~\ref{fig:fig1p5}a of the main text. (b) Micrograph detail of qubits Q5 and Q9 shows the placement (red outline) of the Josephson junction relative to the transmon capacitor and ground plane.}
\end{figure}

The qubit array (Fig.~\ref{fig:fig_device}) has 10 fixed-frequency (single-junction) transmon qubits.
The transmon circuits have a single-ended capacitance to an aluminum ground plane patterned on the \mbox{$\langle 001\rangle$} plane of a double-sided polished silicon substrate.
The intrinsic silicon substrate has dimensions \mbox{$0.35\,\mathrm{mm}\!\times\! 5\,\mathrm{mm}\!\times\!5\,\mathrm{mm}$} along the $x$, $y$, and $z$ axes respectively.
The coplanar waveguide (CPW) geometry of the feedline and readout resonators have a nominal \mbox{$10\,\mu\mathrm{m}$} width and \mbox{$6\,\mu\mathrm{m}$} gap.
The qubit array has 10 charge lines \mbox{(CPW width/gap: $5\,\mu\mathrm{m}/3\,\mu\mathrm{m}$)} for microwave driving of each qubit, although these were not used for the experiment reported here.
Air-bridge crossovers span across the CPWs of the feedline, readout resonators, and charge lines~\cite{rose20}.
The aluminum crossovers have a nominal thickness of 700 nm.

\begin{figure}[htbp]
\includegraphics{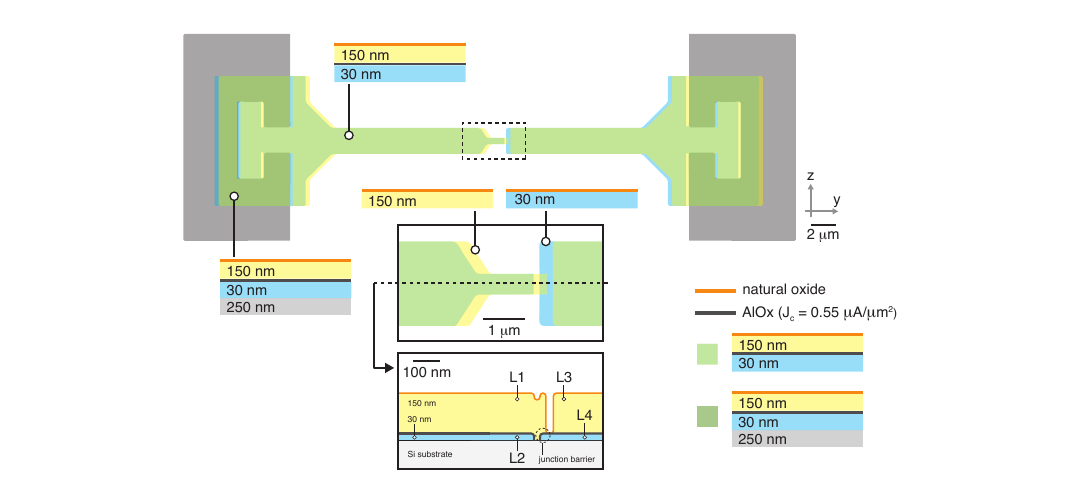}
\caption{\label{fig:jj_diagram}
\textbf{Diagram of the Dolan bridge Josephson junction thin films.}
The Josephson junction was formed by double-angle deposition of aluminum thin films (30-nm followed by 150-nm).
The deposited thin films connect to the 250-nm aluminum (gray) transmon island and ground plane,
which are respectively the left- and right-hand (right- and left-hand) gray regions of the diagram for
\JJplacement{a} (\JJplacement{b}).
The labels \film{1}-4 represent four separate regions of the layered thin films.
\film{1} is a 150-nm film connected to the 250-nm film on the left-hand side and is one of the qubit junction electrodes (as it connects to \film{4} through the junction barrier).
\film{3} is another 150-nm region, which contacts \film{4} and the 250-nm film on the right-hand side.
\film{4} is a 30-nm film that forms the other junction electrode, on the right-hand side of the junction barrier.
}
\end{figure}

The quarter-wavelength fundamental mode of each CPW readout resonator is inductively coupled to the microwave feedline while the opposite end of the resonator is capacitively coupled to its respective transmon capacitor. 
Each transmon capacitor is cross shaped, with the cross arms each having \mbox{$30\,\mu\mathrm{m}$}-width and \mbox{$480\,\mu\mathrm{m}$}-length~(Fig.~\ref{fig:fig_device}b).
Each capacitor is separated from the ground plane by a \mbox{$30\text{-}\mu\mathrm{m}$} gap around the cross.
The transmon qubit capacitors are each separated by $0.85\,\mathrm{mm}$ on center within each row of the qubit array and $2.54\,\mathrm{mm}$ between rows.
The bottom row of transmons are offset to the right (Fig.~\ref{fig:fig_device}a, along the $y$-axis) by $0.225\,\mathrm{mm}$ relative to the top row.
All thin film aluminum, except for the Josephson junction electrodes and air-bridge crossovers, were deposited with a thickness of \mbox{$250\,\mathrm{nm}$}.
After an ion milling procedure, the aluminum Josephson junction electrodes were deposited with thicknesses of \mbox{$30\,\mathrm{nm}$} and \mbox{$150\,\mathrm{nm}$} to create a Dolan-style junction geometry (Fig.~\ref{fig:jj_diagram}).
These thin film thicknesses were measured by atomic force microscopy on the junction electrodes of Q1 and Q3.

This particular qubit array was designed such that each qubit's Josephson junction is either on the right-side (\JJplacement{a}) or left-side (\JJplacement{b}) of the transmon capacitor island (Fig.~\ref{fig:fig_device}b and Table~\ref{tab:qb_parameters}).
The junction placement was intentionally disordered among the qubits in the array
to clarify the direct correspondence between each qubit relaxation rate recovery time scale and junction placement, rather than other, possibly spatially-dependent, differences among the qubits. 
Since the junctions were fabricated from a Dolan bridge double-angle evaporation process there is an inherent asymmetry of the two junction electrodes (Fig.~\ref{fig:jj_diagram}).
The transmon qubits with \JJplacement{a} have the transmon island (ground plane) connected to the thin films on the left-hand (right-hand) side, while qubits with \JJplacement{b} have the opposite orientation.
Consequently, qubits with \JJplacement{a} (\JJplacement{b}) have the 150-nm junction electrode \film{1} connected to the transmon island (ground plane).

The aluminum film thickness differences likely result in superconducting energy gap differences among thin film regions, such as across the qubit Josephson junction~\cite{marc22}, and between the 30-nm and 150-nm thin films and the 250-nm ground plane (and also the transmon island).
Based on superconducting transition temperature $T_c$ measurements of similar films,
we estimate the superconducting gap energy differences between the 150-nm (\film{1}, \film{3}) and 30-nm (\film{2}, \film{4}) films are \mbox{$\approx h(2.8\,\mathrm{GHz})$}, where $h$ is the Planck constant.
Additionally, we estimate the gap difference between the 150-nm films (\film{1} and \film{3}) and the 250-nm aluminum is \mbox{$\approx h(1\,\mathrm{GHz})$}.

The qubit array was placed in a gold-plated copper package.
The outer dimensions of the package are \mbox{$13.0\,\mathrm{mm} \times 52.2\,\mathrm{mm} \times 52.2\,\mathrm{mm}$} along the axes \mbox{$x, y, z$} respectively.
The bottom corners of the qubit array substrate rested upon the gold-plated copper base of the package~\cite{lien19}.
No glue, epoxy, or varnish was used to secure or thermalize the qubit array to the package~\cite{card21, anth22}.
The electrical traces and ground plane of the qubit array are wirebonded (Al-Si alloy wirebonds) to the electrical traces and ground plane of a printed circuit board (PCB), as shown in Figure~\ref{fig:fig_device}~\cite{vent96}.
The PCB dielectric is \mbox{Rogers TMM10} material.
We note that the PCB dielectric could be a contributing source of ionizing radiation by 
containing trace quantities of \mbox{potassium-40} and progeny nuclei of the uranium and thorium decay chains~\cite{card23, loer24}.

The critical current density of each Josephson junction \mbox{($J_c = 0.55\,\mu\mathrm{A}/\mu\mathrm{m}^2$)} is estimated based on room-temperature conductance measurements of witness junctions from the same wafer.
The anharmonicity of each transmon is \mbox{$160 \pm 5\,\mathrm{MHz}$}, where the uncertainty reflects variation among qubits in a representative qubit array with the same capacitor and readout coupling geometry.

Qubit decay rates were measured periodically after each collection of 100 entries ($\approx 45\,\mathrm{min}$ real time) using an inversion-recovery pulse sequence as is traditionally performed to characterize a superconducting qubit's $1/T_1$ energy-decay rate~\cite{kran19}.
In Table~\ref{tab:qb_parameters}, we report the median decay rate for each qubit as measured through \run{9} to \run{13}, and additional measurements thereafter.
The upper and lower uncertainty represents deviation from the median for the 84.1$^\text{st}$ and 15.9$^\text{th}$ percentile value (which corresponds to $\pm 1$ standard deviation for a Gaussian distribution).
The delay duration between preparation and measurement in each pulse sequence was varied from \mbox{$8\,\mathrm{ns}$} to \mbox{$117.608\,\mu\mathrm{s}$} in 50 equal steps. 
The full range of delay times were repeated for 500 single-shots per delay duration, with \mbox{$600\,\mu\mathrm{s}$} between each single-shot readout and the following preparation pulse.
An exponential decay rate was extracted from a least squares fit to the averaged single-shot voltages.

\subsubsection{Qubit relaxation rate recovery}\label{sec:jjorientation}
The direct correspondence between the qubit recovery timescale after a \stc{} event and junction placement (Table~\ref{tab:qb_parameters})
could be explained by the relative rates that quasiparticles tunnel into the thin films of the junction electrodes (\film{1} and \film{4}).
The different recovery timescales may result from the combined influence of
\mbox{1. a significant} difference of quasiparticle density between each transmon island and the ground plane, as well as
\mbox{2. the junction} orientation with respect to the 30-nm junction electrode
(\film{4}, having the highest superconducting gap energy) connection to either the transmon island or ground plane.

For example, quasiparticles may predominantly tunnel from the transmon island into the nearest \mbox{150-nm} film (which is either \film{1} or \film{3} depending on \JJplacement{a} or \JJplacement{b}, respectively) if there is significantly stronger trapping and/or recombination in the ground plane compared to the transmon islands. We note that the air-bridge crossovers may contribute to quasiparticle trapping in the ground plane.
In the case of \JJplacement{a}, quasiparticles that tunnel from the transmon island and into the 150-nm junction electrode \film{1}~(Fig.~\ref{fig:jj_diagram}) can also tunnel across the Josephson junction and cause qubit relaxation.
Excess quasiparticle-induced qubit relaxation would persist on the timescale that there are quasiparticles in the 150-nm junction electrode \film{1}.
However, in the case of \JJplacement{b}, quasiparticles, again sourced the transmon island,  would tunnel into the 150-nm thin film \film{3}~(Fig.~\ref{fig:jj_diagram}) and are prevented from tunneling across the qubit junction by the higher-gap film \film{4}~\cite{diam22}.
In this case of \JJplacement{b}, quasiparticle-induced qubit relaxation requires that quasiparticles from the transmon island first enter \film{4}, which is expected to be suppressed by an Arrhenius factor \mbox{$\sim e^{-\delta\Delta/k_\mathrm{B}T}$}, where \mbox{$\delta\Delta\approx h(2.8\,\mathrm{GHz})$} is the expected superconducting gap energy difference between these films and \mbox{$k_\mathrm{B}T$} is the effective energy of the quasiparticles~\cite{conn23}.

\begin{table}[htbp]
\caption{\label{tab:qb_parameters}
\textbf{Qubit array parameters.}
The column ``JJ placement'' indicates the location of the Dolan bridge Josephson junction relative to transmon capacitor and ground plane (see Figure~\ref{fig:fig_device}); ``\JJplacementLetterOnly{a}'' indicates that the 30-nm electrode \film{4} is on the ground-plane-side of the junction, whereas ``\JJplacementLetterOnly{b}'' indicates that the 150-nm electrode \film{1} is on the ground-plane-side of the junction.
The ``Recovery'' column reports the time-constant of decay rate dynamics averaged over all events.
}
\begin{tabular}{lcclcc}
\hline
& Qubit & Readout & \multicolumn{1}{c}{$T_1 = 1/\Gamma_1$} & Junction & Recovery \\
& Frequency (GHz) & Frequency (GHz) & \multicolumn{1}{c}{$(\mu\mathrm{s})$} & placement & (ms)\\ \cline{1-6}
{Q1}	&	4.534	&	6.759	&	\hspace{0.25cm} $53\substack{+9\\ -10}$	&	\JJplacementLetterOnly{a}	 & 5.9\\
{Q2}	&	4.370	&	6.648	&	\hspace{0.25cm} $49\substack{+8\\ -14}$	&	\JJplacementLetterOnly{a}	& 6.6\\
{Q3}	&	4.949	&	6.889	&	\hspace{0.25cm} $42\substack{+9\\ -9}$	&	\JJplacementLetterOnly{b}	& 0.8\\
{Q4}	&	4.697	&	6.789	&	\hspace{0.25cm} $16\substack{+1\\ -2}$	&	\JJplacementLetterOnly{a}	& 6.5\\
{Q5}	&	4.453	&	6.693	&	\hspace{0.25cm} $40\substack{+6\\ -6}$	&	\JJplacementLetterOnly{a}	& 6.0\\
{Q6}	&	5.015	&	6.926	&	\hspace{0.25cm} $43\substack{+7\\ -8}$	&	\JJplacementLetterOnly{b}	& 0.8\\
{Q7}	&	4.840	&	6.825	&	\hspace{0.25cm} $57\substack{+16\\ -17}$	&	\JJplacementLetterOnly{b}	& 0.7\\
{Q8}	&	4.501	&	6.729	&	\hspace{0.25cm} $55\substack{+11\\ -16}$	&	\JJplacementLetterOnly{a}	& 6.5\\
{Q9}	&	5.155	&	6.960	&	\hspace{0.25cm} $69\substack{+11\\ -11}$	&	\JJplacementLetterOnly{b}	& 0.7\\
{Q10}	&	4.916	&	6.853	&	\hspace{0.25cm} $47\substack{+8\\ -9}$	&	\JJplacementLetterOnly{b}	& 0.8\\
\end{tabular}
\end{table}

\subsubsection{Scintillating detectors}\label{sec:detectors}
\paragraph{Detector construction}
Each detector is a rectangular prism of scintillating polymer wrapped with reflective film and optically coupled to a photomultiplier tube (PMT) that produces pulses of current that are commensurate with the energy deposited in the scintillator from ionizing radiation impacts. 
The materials and components of each scintillator are provided in Table~\ref{tab:detectors}.
The scintillating materials (EJ-200 and BC-412) were chosen for their long light attenuation lengths (\mbox{$>2\,\mathrm{m}$}), which is relevant to ensure the PMT efficiently collects light produced anywhere within the scintillator volume.
The detectors are labeled alphabetically and are grouped in three pairs of similar construction
\mbox{(\det{a} \& \det{b}, \det{c} \& \det{d}, \det{e} \& \det{f})}.
\begin{table}[htbp]
\centering
\caption{
\textbf{Summary of the scintillating detectors.}
The scintillating material of each detector is optically coupled to a PMT, operated with a high-voltage bias. The volumetric center position of each detector is reported relative to the volumetric center of the qubit array substrate. The detector dimensions and position coordinates are ordered as ($x$, $y$, $z$). 
}
\label{tab:detectors}
\begin{tabular}{cr@{\hskip 10pt}r@{\hskip 10pt}c r@{$\times$}c@{$\times$}l r@{, }c@{, }l}
\hline

Detector & Material & \multicolumn{1}{c}{PMT} & \multicolumn{1}{c}{HV bias (kV)} & \multicolumn{3}{c}{Dimensions (cm)} & \multicolumn{3}{c}{Position (cm)} \\ \hline
A & EJ-200 & Hamamatsu R9800 & -1.3 & 51.0 & 7.2 & 2.0 & (-13.07 & 6.59 & -43.77) \\
B & EJ-200 & Hamamatsu R9800 & -1.3 & 51.0 & 7.2 & 2.0 & (-13.07 & -0.61 & -44.06) \\
C & EJ-200 & Hamamatsu R7724 & -1.2 & 60.0 & 7.0 & 7.0 & (-8.57 & -0.61 & -48.88) \\
D & EJ-200 & Hamamatsu R7724 & -1.2 & 60.0 & 7.0 & 7.0 & (-8.57 & -0.61 & -56.51) \\
E & BC-412 & ET Enterprises 9266 & -1.0 & 100.0 & 5.72 & 6.0 & (-5.67 & 5.75 & -49.51) \\
F & BC-412 & ET Enterprises 9266 & -1.0 & 100.0 & 5.72 & 6.0 & (-5.67 & 5.75 & -57.01) \\ \hline
\end{tabular}
\end{table}

\paragraph{Detector readout}
The use of the scintillating detectors for measurements of cosmic rays required the discrimination, filtering, and digitization of pulse signals produced by the scintillators and PMTs.
Since the PMT produces rapid (\mbox{$\approx 10\,\mathrm{ns}$}) pulses of current, we used a charge pre-amplifier as a low-pass integrator of the PMT current.
The pre-amplifier effectively accumulates charge on a capacitor, which is then measured as a voltage pulse with the analog-to-digital converter.
The amplitude of each voltage pulse is thus proportional to the number of photons collected by PMT from the scintillator, which itself is proportional to the total deposited energy in the scintillator from an impact of ionizing radiation.
We operated each PMT using a negative high-voltage (HV) bias, chosen for sufficiently high detection efficiency of cosmic rays while minimizing dark counts.

Each pre-amplified detector signal was fed into a commercial FPGA-based analog-to-digital converter (Caen~DT5725S), which has a \mbox{$250\,\text{MS/s}$} sampling rate and a \mbox{14-bit} amplitude resolution. 
The on-board FPGA was programmed to discriminate, filter, and digitize each pulse timestamp and amplitude.
The recorded pulse amplitude and arrival time were stored offline for further filtering and analysis.
Table~\ref{tab:det_equipment} lists the electronic equipment used for detector operation and data collection.
\begin{table}[htbp]
    \centering
    \caption{\textbf{Scintillating detector readout equipment.}}
    \label{tab:det_equipment}
    \begin{tabular}{l@{\hskip 10pt}l@{\hskip 10pt}l}
        \hline
		\multicolumn{1}{c}{Component} & \multicolumn{1}{c}{Manufacturer} & \multicolumn{1}{c}{Model} \\ \hline
		Charge pre-amplifier & Cremat & CR-113-R2.1 \& CR-150-R6\\
		Analog-to-digital converter & Caen & DT5725S (\mbox{14-bit}, \mbox{$250\,\text{MS/s}$})\\
		HV Supply & Caen & DT5533EN\\ \hline
    \end{tabular}
\end{table} 

\subsubsection{Positions of the detectors and qubit array}\label{sec:positions}
The orientation and aspect ratio of the detectors and qubit array substrate influenced the rate and distribution of energy depositions from cosmogenic particles.
Additionally, the positioning of detectors determined the rate of coincidence events among the detectors themselves and the coincidences with the qubit array.
We chose a geometric configuration of the detectors that enabled an accurate calibration of our cosmic-ray model (detector response and cosmic-ray flux)
and a sufficient rate of qubit-detector coincidences.

The qubit array was mounted vertically in the cryostat such that the normal vector of the thin film ($x$-axis) points to the horizon (Fig.~\ref{fig:fig1p5}a).
We note that a rotation of the chip from a vertical to horizontal orientation would approximately double the cosmic-ray impact rate.
A vertical orientation exposes the qubit array to cosmic rays that can traverse long distances \mbox{$\sim\!5\,\mathrm{mm}$} through the substrate, thus depositing higher energy on average compared to a horizontal orientation.

We placed the scintillating detectors outside the experiment cryostat and underneath the qubit array as shown in Figure~\ref{fig:fig1p0}, Figure~\ref{fig:fig3}a, and Figure~\ref{fig:g4_scene}.
An accurate representation of the experiment's geometric configuration was required for modeling interaction cross-sections from cosmic rays using \geant{} (Section~\ref{sec:g4}).
The measured positions of the detectors, relative to the qubit array, are included in Table~\ref{tab:detectors}.
The detectors were placed and aligned according to reference coordinates marked on the floor of the laboratory.
We defined the coordinate system relative to the volumetric center of the qubit array substrate when the cryostat cans were removed prior to cooldown.
Distance measurements were performed using a laser range finder, meter stick, and calipers.
The alignment of each detector was aided by a laser level and plumb bob.
The uncertainty of the detector positions relative to each other is \mbox{$\approx 2\,\mathrm{mm}$} and the
uncertainty of the qubit array position (relative to all detectors) is \mbox{$\approx 5\,\mathrm{mm}$}.
This geometric uncertainty does not significantly affect our estimation of interaction cross-sections.
For the chosen detector arrangement, we found the calculated interaction cross-section are insensitive to deviations of position that are small compared to the detector dimensions.

The detectors were arranged directly below the qubit array, nearly centered on the $z$-axis, to maximize the qubit-detector coincidence rate.
The qubit array was mostly exposed to cosmic rays that originated from near-zenith and azimuthal angles along the $x$-axis, symmetrically entering from the front-side and back-side of the substrate chip.
Accordingly, we aligned the longest dimension of each detector with the $x$-axis to maximize coincidences from these azimuthal angles.

We created two vertical stacks of three scintillators each.
Each detector stack provided a sufficient rate of two-fold and three-fold detector-detector coincidences which enabled the calibration of the detector energy response (Section~\ref{sec:det_cal}).
We placed the two stacks adjacent to each other (along the $y$-axis).
Lead bricks were arranged around the stacks to reduce the flux of gamma radiation incident on the detectors.

\subsubsection{Qubit measurement pulse sequence}\label{sec:pulse_sequence}
\begin{figure}[htbp]
\includegraphics{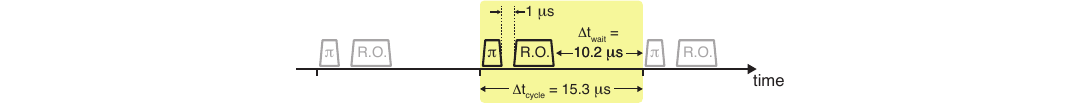}
\caption{\label{fig:msmt_cycle}
\textbf{Qubit control and readout pulse sequence.}
The pulse sequence of a single measurement cycle \mbox{($\dtcycle=\valueCycleDurationAvgApprox{}\,\mu\mathrm{s}$)} included
a qubit control \mbox{$\pi$-pulse},
a \mbox{$1\text{-}\mu\mathrm{s}$} delay for qubit decay,
and a \mbox{$4\text{-}\mu\mathrm{s}$} readout pulse for single-shot state discrimination, which was followed by a wait-time duration of \mbox{$\dtdelay=10.2\,\mu\mathrm{s}$} before the next measurement cycle.
This pulse sequence was frequency-multiplexed to address qubits or resonators simultaneously.
}
\end{figure}

Qubit measurements for all data runs (Section~\ref{sec:summary_runs}) were performed by repeated cycles of a pulse sequence for qubit control and single-shot readout.
Each measurement cycle has a duration period of $\dtcycle=\valueCycleDurationAvgApprox{}\,\mu\mathrm{s}$.
The pulse sequence (Fig.~\ref{fig:msmt_cycle}) performed during each cycle consists of
a control pulse for each qubit \mbox{($\pi$-pulse, 100 ns)},
a \mbox{$1\text{-}\mu\mathrm{s}$} delay,
a single-shot readout ($4\,\mu\mathrm{s}$) for each qubit,
and a wait-time (\mbox{$\dtdelay=10.2\,\mu\mathrm{s}$}) before the following measurement cycle.
The wait-time following readout was chosen for sufficient resonator ring-down and to have a data collection time \mbox{($\approx 15\,\mathrm{s}$)} per entry that was comparable to the downtime \mbox{($\approx 12\,\mathrm{s}$)} between data entries (Section~\ref{sec:summary_runs}).

We used frequency multiplexed microwave pulses for qubit control and resonator readout.
The readout pulse amplitudes and frequencies were individually set to maximize the single-shot fidelity of separation between qubit pointer states while minimizing readout-induced excitations of the transmon circuit outside the qubit manifold~\cite{sank16, khez23}.

An instance of qubit relaxation occurred whenever a qubit was prepared in the excited state after the control pulse and, following the \mbox{$1\text{-}\mu\mathrm{s}$} delay, it was then found in the ground state.
While readout informed of qubit decay, it also informed the preparation state for the next measurement cycle.
All instances of measured qubit decay occurred from post-selected measurement cycles, which were conditioned on a ground-state result from the readout of the previous measurement cycle.
If a qubit was found in the ground state, the qubit remained in the ground state with high probability after the single-shot measurement and throughout the $\dtdelay$ wait-time before the next measurement cycle.
Since the qubit remained in the ground-state during the wait-time, the control pulse transitioned the qubit to the excited state (assuming perfect $\pi$-pulse fidelity).

We checked that qubit excitation was unlikely by measuring excitation rates using an adapted version of the presented pulse sequence; every other cycle presented in Figure~\ref{fig:msmt_cycle} was replaced with a sequence lacking a $\pi$-pulse, which allowed for excitation measurements interleaved with \stc{} relaxation event detection.
We found that qubit excitation rarely occurred (\mbox{$\Gamma_\uparrow \ll 1/\dtdelay$}) both in steady-state and during \stc{} relaxation events.

The pulse sequence allowed for detection of qubit relaxation not only during the \mbox{$1\text{-}\mu\mathrm{s}$} delay, but also during the single-shot readout process.
We estimated an effective delay duration of \mbox{$\Delta t = 3\,\mu\mathrm{s}$} by comparing the traditional $1/T_1$ energy-decay measurements (Table~\ref{tab:qb_parameters}) to the steady-state decay probability measured by this pulse sequence~\mbox{(Fig.~\ref{fig:msmt_cycle})}.

Excited-state readout results are not highly informative of the qubit state after the $\dtdelay$ wait-time due to the effect of qubit energy-relaxation (\mbox{$\Gamma_\downarrow \sim 1/\dtdelay$}), where $\Gamma_\downarrow$ is rate of energy-relaxation of the qubit.
A measurement result that occurs in a cycle after an excited-state readout of the previous cycle does not inform if qubit relaxation occurred and the result is used only to condition the next measurement cycle.

\subsubsection{Synchronization of qubit and detector measurements}\label{sec:ref_timing}
This experiment correlated qubit relaxation errors and cosmic ray detection, which relied on the synchronization of qubit and detector data sets.
Throughout data collection, we recorded all detector pulses that occurred while qubits were measured.
We synchronized detector and qubit measurement data by referencing each detector pulse arrival to a specific single-shot qubit measurement.

We synchronized the detector and qubit data by producing a square reference pulse from the qubit measurement hardware and recording its occurrence with an additional channel of the analog-to-digital (ADC) converter for the detectors.
Within each entry of $10^6$ qubit measurement cycles, a reference pulse was generated by the \mbox{Quantum Machines OPX+} after the first 100 qubit measurement cycles and every 100 cycles thereafter.
After data collection, we found all timestamps of the reference pulses in the detector data, as recorded by the ADC with 4-ns resolution.
A group of 10,000 consecutive reference pulses marked the $10^6$ measurement cycles of each data entry.
Accordingly, we assigned each reference-pulse timestamp to its respective measurement cycle index within an entry.
For each entry, we performed a linear interpolation between reference-pulse timestamps versus cycle index, which served to relate every scintillator pulse timestamp to its contemporary measurement cycle index within the entry.

The assignment of each detector pulse to a measurement cycle coarse-grained their \mbox{4-ns}-resolution arrival times to \mbox{$\dtcycle=\valueCycleDurationAvgApprox{}\,\mu\mathrm{s}$} time bins.
The coarse graining procedure was unlikely to separate the pulses of detector-detector cosmic-ray coincidences into consecutive cycles (which would result in false-negatives)
because detector-detector coincidences occur within a short duration (\mbox{$\approx 100\,\mathrm{ns}$}, observed in the 4-ns resolution detector data) compared to the \mbox{$\valueCycleDurationAvgApprox{}\text{-}\mu\mathrm{s}$} measurement cycle window.

\subsection{Identification of Spatiotemporally Correlated Qubit Relaxation Events}\label{sec:mf}
We consider \stc{} error events that have time scales and spatial scales that could affect the performance of quantum error correction codes, and we define \stc{} qubit relaxation events in the context of the multi-qubit correlated energy-relaxation we observe in our qubit array device-under-test.
Throughout this work we refer to \stc{} error events which have \mbox{$\mathcal{O}(1\,\mathrm{ms})$} recovery timescales that could be commensurate to many rounds of quantum error correction and \mbox{$\mathcal{O}(1\,\mathrm{mm})$} length scales that might challenge the operation of error decoding and correction protocols.
We observe \stc{} qubit relaxation events that have \mbox{$\mathcal{O}(1\,\mathrm{ms})$} timescales and that affect the relaxation rates of multiple qubits.
These events occur in great excess compared to an expectation from an independent error model and each qubit's steady-state rate of energy-relaxation (Table~\ref{tab:qb_parameters}).

Qubit energy-relaxation is the primary mechanism of the \stc{} errors observed in our device.
We identified such error events by repetitively monitoring for instances of qubit relaxation (Section~\ref{sec:pulse_sequence}).
An event is defined operationally by a cross-correlation filtering procedure~\cite{toom89} that we applied to time-series data of monitored qubit relaxation.

\begin{figure}[htbp]
\includegraphics[width=\textwidth]{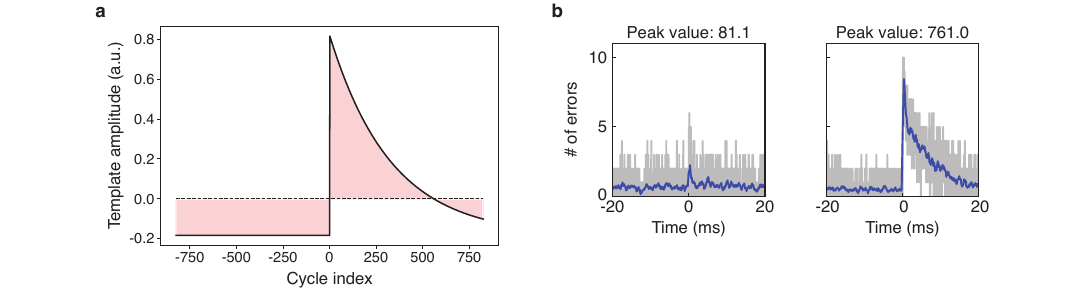}
\caption{\label{fig:mf_template}
\textbf{Spatiotemporally correlated relaxation event identification.}
(a) The cross-correlation filter template is a one-sided exponential decay. The filter amplitude is offset to have a zero-sum integral value.
(b) Example data for selected values of cross-correlation peaks.
The number of simultaneous energy-relaxation errors is shown as raw counts (gray) and its moving average per \mbox{40 cycles} (blue).
}
\end{figure}

We defined a filter template (Fig.~\ref{fig:mf_template}a) for the expected dynamics of qubit energy-relaxation rates during an event.
The template captures the characteristic temporal behavior of the spatially-correlated energy-relaxation observed in the data, as it resembles the rapid onset and \mbox{$\mathcal{O}(1\,\mathrm{ms})$} recovery dynamics of excess qubit relaxation during an event.
We defined the template as a \mbox{$25\,\mathrm{ms}$} time-series (1,648 cycles), for which the initial \mbox{$12.5\,\mathrm{ms}$} has no change in value while the following \mbox{$12.5\,\mathrm{ms}$} is an exponential decay with a \mbox{$5\,\mathrm{ms}$} recovery time-constant~\mbox{(Fig.~\ref{fig:mf_template}a)}.

For each of the \valueNentries{} data entries (\mbox{$10^6\,\mathrm{cycles}$} per entry) of the experiment, we identified \stc{} qubit relaxation events using the following procedure:
\begin{enumerate}
\item Evaluate if qubit relaxation was detected, as \verb|True|/\verb|False|, for each measurement cycle for each qubit.
Qubit relaxation in a given cycle is registered as \verb|True| if the qubit is measured in the ground state (in the cycle under evaluation) and it was measured in the ground state during the previous cycle (Section~\ref{sec:pulse_sequence}).
\item Evaluate the total counts of qubit relaxation per measurement cycle by summing over all qubits within a cycle. For each entry, this produces a time-series array of integers ranging from zero to the total number of qubits.
\item Calculate the cross-correlation between the time-series data and the filter template.
The cross-correlation was evaluated by discrete convolution between the data and the time-reverse of the template.
Before convolution, the template was offset to have zero-mean which removed bias on the cross-correlation due to the total qubit relaxation on time scales greater than \mbox{$25\,\mathrm{ms}$}.
\item Find event candidates by peak-finding with the Python function \verb|scipy.signal.find_peaks|~\cite{virt20}.
The measurement cycle index of each identified peak marks the onset of a \stc{} relaxation event.
When we performed cross-correlation peak identification, we required the duration between peaks to exceed \mbox{$12.5\,\mathrm{ms}$} (at least half the duration of the cross-correlation filter template).
This ensured that multiple peaks were not identified within a single event candidate.
An event candidate was required to have a cross-correlation peak value above {50}.
This threshold value was chosen to ensure high acceptance of 
true-positive qubit events.
\item Register events from the candidates if a cross-correlation peak value exceeds {105}. This final filtering threshold was chosen to limit false-positive events.
\end{enumerate}

We recorded each cross-correlation peak that exceeds a threshold value of 50, 
which were identified as event candidates (Step \#4) before further filtering for event acceptance.
The event candidate threshold of 50 is lower than the event threshold.
Since the quasi-steady-state, or baseline, decay rate of each qubit fluctuated over the duration of the experiment (Table~\ref{tab:qb_parameters}),
the event candidate detection rate was influenced by background $1/T_1$ fluctuations of individual qubits throughout the duration of the experiment.
We chose an event acceptance threshold value of 105 to maintain a low rate of false-positive detection of the events we study in this experiment.
\begin{figure}[htbp]
\includegraphics[width=\textwidth]{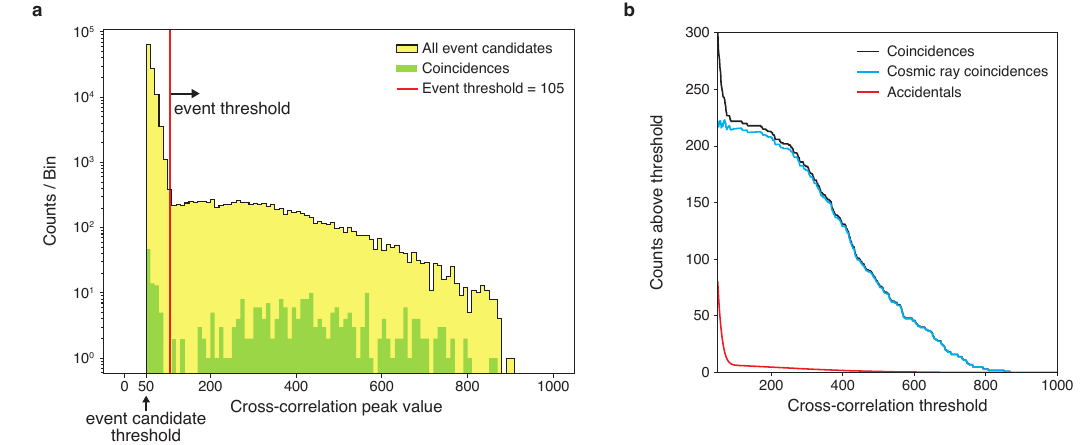}
\caption{\label{fig:mf_height}
\textbf{Cross-correlation peak distribution.}
(a) A histogram of all identified cross-correlation peak values for candidate events (yellow) is shown with
the subset of candidate events that are coincident with a detector pulse (green).
We identified \stc{} events as the candidates that have a peak value above 105 (red line).
(b) Qubit-detector coincidence counts versus the cross-correlation threshold.
The estimated accidental counts (red) are subtracted from the observed coincidences (black) to give the cosmic-ray coincidence counts (blue)~\mbox{(Eq.~\ref{eq:rateQD})}.
These data, considered with the background model, support the claim that most coincidence event candidates below threshold are likely accidentals~(false-positives).
}
\end{figure}

The event threshold value of 105 was determined as the lowest threshold that also minimized false-positive event identification.
We limited false-positive event assignment by ensuring that the inter-arrival distribution of \stc{} events had acceptable goodness-of-fit to an exponential distribution.
This criteria was chosen because we consider that true-positive \stc{} events should not be correlated to each other. For instance, an energy deposition from an ionizing radiation particle would not affect the likelihood of another particle impact at a much later time (\mbox{$>12.5\,\mathrm{ms}$}).

Independent $1/T_1$ decay rate fluctuations of each qubit can result in spurious (random) spatial correlations.
Decay rate fluctuations are ubiquitously observed in superconducting circuit devices, generally from intermittent interactions with and among two-level-systems~\cite{harr17, burn19, schl19, carr22}.
The spectral density of decay-rate fluctuations follows a power-law ($1/f^\alpha$) distribution. 
Random, but significant, spatial correlations can arise from the spurious correlation of independent power-law noise processes. 

Figure~\ref{fig:mf_height}a shows a histogram of all candidate cross-correlation peaks (blue) identified in the analysis.
\mbox{Low-valued (\mbox{$\lesssim 105$})} cross-correlation event candidates occurred more frequently compared to the accepted events above threshold.
The peaks that occurred in coincidence with a detector (green) show distinct populations among low and high cross-correlation peak values.
Most event candidates below threshold are likely false-positive events, as evidenced by the proportionality of coincidence and accidental counts (Section~\ref{sec:interarrival_bkgnd}) at low cross-correlation thresholds~\mbox{(Fig.~\ref{fig:mf_height}b)}.

Additionally, we found \stc{} events occurred more frequently when the qubit array was exposed to heightened levels of gamma-ray radiation (Section~\ref{sec:cs137}), which established that the \stc{} events, as defined, can occur from ionizing radiation energy depositions from a non-cosmogenic source.

\subsection{Summary of Data Runs}\label{sec:summary_runs}
\begin{table}[htbp]
\caption{\label{tab:runs}
\textbf{Summary of Data Runs.}
We analyzed \valueNentries{} entries (excludes \run{7}) for the experiment analysis giving the event rate $\rateQ$ and all qubit-detector coincidences.
$^*$These data were analyzed for calibration of detector efficiency and muon flux~\mbox{(Section~\ref{sec:eff_flux})}.
Detector data, that was collected during, between and after \run{0} entries, were analyzed for detector energy-response calibration~\mbox{(Section~\ref{sec:det_cal_subsec})}.
$^{**}$Comparison run with a \Cs{} gamma-ray source~\mbox{(Section~\ref{sec:cs137})}.
}
\begin{tabular}{rrcr}
\hline
\multicolumn{1}{c}{\multirow{2}{*}{Run}} & \multicolumn{1}{c}{\multirow{2}{*}{Entries}} & \multirow{2}{1.5cm}{\centering \# of Qubits} & \multirow{2}{2cm}{\centering Inverse Event Rate (s)} \\
& & & \\ \hline
\multicolumn{1}{r}{$^*$\run{0}} & 2,627 & 8 & $115$ \hspace{0.75cm} \\
\multicolumn{1}{r}{\run{1}} & 68 & 8 & $79$ \hspace{0.75cm} \\
\multicolumn{1}{r}{\run{2}} & 2,785 & 8 & $102$ \hspace{0.75cm} \\
\multicolumn{1}{r}{\run{3}} & 913 & 8 & $92$ \hspace{0.75cm} \\
\multicolumn{1}{r}{\run{4}} & 5,458 & 8 & $103$ \hspace{0.75cm} \\
\multicolumn{1}{r}{\run{5}} & 8,158 & 8 & $101$ \hspace{0.75cm} \\
\multicolumn{1}{r}{\run{6}} & 4,118 & 8 & $103$ \hspace{0.75cm} \\
\multicolumn{1}{r}{$^{**}$\run{7}} & 3,878 & 8 & $10.2$ \hspace{0.50cm} \\
\multicolumn{1}{r}{\run{8}} & 4,227 & 8 & $100$ \hspace{0.75cm} \\
\multicolumn{1}{r}{\run{9}} & 8,781 & 10 & $101$ \hspace{0.75cm} \\
\multicolumn{1}{r}{\run{10}} & 2,267 & 10 & $98$ \hspace{0.75cm} \\
\multicolumn{1}{r}{\run{11}} & 1,945 & 10 & $107$ \hspace{0.75cm} \\
\multicolumn{1}{r}{\run{12}} & 2,383 & 10 & $95$ \hspace{0.75cm} \\
\multicolumn{1}{r}{\run{13}} & 19,090 & 10 & $101$ \hspace{0.75cm} \\ \hline
\end{tabular}
\end{table}

The experiment data collected was divided over multiple runs, each containing many repeated data entries.
Each data entry is a continuous time stream of \mbox{$10^6$} consecutive qubit measurements \mbox{(Section~\ref{sec:pulse_sequence})}, amounting to approximately \mbox{$15\,\mathrm{s}$} per entry.
During each run, there was a \mbox{$12\,\mathrm{s}$} period between each entry when the data collection was inactive due to the computational time required to retrieve all single-shot qubit measurements from hardware and save to disk.

The qubits {Q6} and {Q9} were omitted from measurements for \run{0} through \run{8} due to a control programming error.
The inclusion of these qubits (for \run{9} through \run{13}) did not significantly affect the detection rate of spatiotemporally correlated qubit relaxation events (Table~\ref{tab:runs}).
The measurement cycle duration for \run{9} through \run{13} is 1.4\% longer (208 ns, Section~\ref{sec:pulse_sequence}) than the measurement cycle duration for \run{0} through \run{7}.
The cycle duration \mbox{$\valueCycleDurationAvg{}\,\mu\mathrm{s}$} (averaged over entries) was used for the rate calculations that depend on an inter-arrival duration (such as coincidence and accidental rates) since all entries were aggregated for the coincidence analysis.
The systematic error due to this 1.4\% difference in cycle duration would have a negligible effect on our results (e.g.~the rate of cosmic-ray-induced qubit events) since the identification efficiency of coincidences is insensitive to this small change of the coincidence window duration.

\subsubsection{Qubit-detector coincidences}\label{sec:qb_det_coin}
Qubit-detector coincidences are defined by the relative timing of a \stc{} qubit relaxation event and its nearest-in-time detector pulse.
Although detector-detector coincidences have a short characteristic inter-arrival delay duration of \mbox{$\approx 100\,\mathrm{ns}$} \mbox{(Section~\ref{sec:ref_timing})},
the qubit-detector coincidences have a longer inter-arrival delay (\mbox{$\approx 50\,\mu\mathrm{s}$}) between a qubit event onset and its nearest-in-time detector pulse.
This resulted from a timing imprecision of the cross-correlation technique (Section~\ref{sec:mf}) due to sampling of qubit relaxation (Section~\ref{sec:pulse_sequence}).
As defined, a coincidence occurs if a detector pulse is in the same, preceding, or following measurement cycle as a \stc{} qubit relaxation event, which corresponds to coincidence observation window of \mbox{$\dtcoin=3\dtcycle$}.
This is equivalent to the inter-arrival delay $\Delta t_k$  satisfying the condition \mbox{$|\Delta t_k| \le \dtcoin/2$}.
Note that the measurement cycle duration $\dtcycle$ sets the resolution of the inter-arrival timing \mbox{(Section~\ref{sec:ref_timing})} and we define the timestamp of each measurement cycle $t_k$ such that $\dtcycle$ spans \mbox{$[t_k-\dtcycle/2, t_k+\dtcycle/2]$}.

\begin{figure}[htbp]
\includegraphics[width=\textwidth]{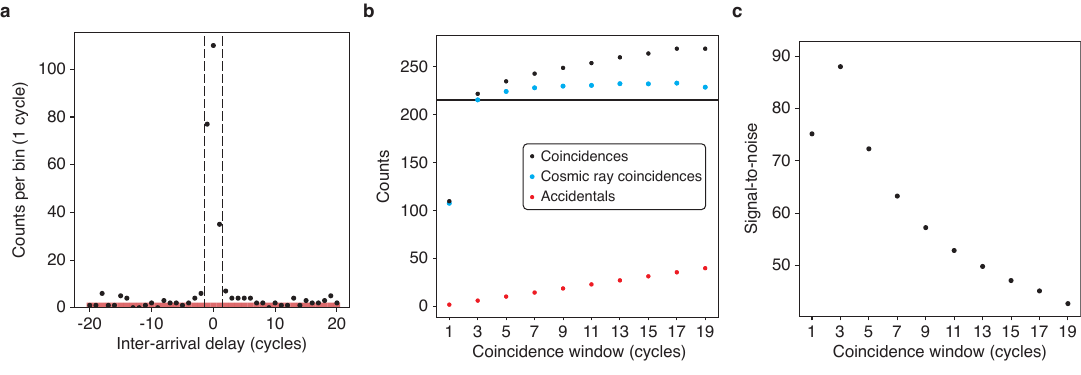}
\caption{\label{fig:qb_det_coin} \textbf{Qubit-detector coincidences versus coincidence window.}
(a) A histogram of the \det{QS} inter-arrival distribution with \mbox{$1\dtcycle$-bin-width}.
The \mbox{$3\dtcycle$-duration} coincidence window is indicated by the interval between the dashed lines.
(b) The total counts of observed coincidences and expected number of accidentals increases with coincidence window duration.
Their difference gives an estimate of the cosmic-ray coincidence counts (Eq.~\ref{eq:rateQD}), which saturates for large coincidence windows.
(c) The statistical signal-to-noise ratio for measured coincidences versus the coincidence window duration.
}
\end{figure}

The definition of the coincidence observation window $\dtcoin$ was informed by both measured and simulated qubit-detector inter-arrival distributions.
Figure~\ref{fig:qb_det_coin}a shows a histogram \mbox{$1\dtcycle$-bin-width} of the measured inter-arrival distribution for each qubit array (\det{q}) event and its nearest-in-time detector pulse (\det{s}) from any of the detectors.
There is a significant excess of inter-arrival counts above the expected background for inter-arrival delays within and immediately outside the center measurement cycle.
Although ionization, phonon transport, and quasiparticle tunneling are expected to occur within a \mbox{$\valueCycleDurationAvgApprox{}\,\mu\mathrm{s}$} measurement cycle~\cite{mart19, mart21},
our time-resolution for event identification was limited to \mbox{$\approx 3\,\mathrm{cycles}$} due to the 
sampling statistics of qubit relaxation.

We assigned an efficiency of \mbox{$\epCoin = \valueEffQD{}$} for coincidence identification within the coincidence observation window \mbox{$\dtcoin=3\dtcycle$} based on both observed and simulated inter-arrival distributions.
We created an expected inter-arrival delay distribution for qubit-detector coincidences by simulating qubit relaxation data (with \mbox{5,000} known \stc{} event arrival times) and performing cross-correlation peak detection (Section~\ref{sec:mf}).
The simulated inter-arrival delays informed that 6\%~of the simulated cosmic-ray coincidences occurred outside the coincidence window, which is consistent with the additional cosmic-ray coincidences observed (\mbox{$6\pm1\%$}) for coincidence window durations \mbox{$\ge 7\dtcycle$}~\mbox{(Fig.~\ref{fig:qb_det_coin}b)}.

We motivate the \mbox{$3\dtcycle$} coincidence window definition by separating the measured coincidences into signal and noise components, which are
cosmic-ray and accidental coincidences, respectively.
We define a statistical signal-to-noise ratio (SNR) as the number of coincidences observed $\NQD$ divided by $\sqrt{\NbkgndQD}$, which corresponds to the expected scale of deviation from the mean accidental counts $\NbkgndQD$.
This SNR quantifies the overall correlation between qubit events and detector pulses.
The SNR scales with the number of standard deviations ($z$-score) that our data deviates from the expected counts if there were no correlation between the qubit events and detectors.
Figure~\ref{fig:qb_det_coin}c shows the statistical signal-to-noise ratio dependence on the coincidence window duration. 
The statistical signal-to-noise ratio is maximized for the \mbox{$3\dtcycle$-window}, as this duration maximizes the acceptance of cosmic-ray coincidences while minimizing accidentals.

\subsubsection{Occurrence rate estimation}\label{sec:rates}
The observed number of qubit relaxation events, pulses from each detector, and their coincidence combinations are described by Poisson random variables.
The treatment of each variable as a Poisson process is consistent with our observation of exponentially distributed inter-arrival distributions for detector pulses and qubit events.
Each Poisson process is characterized by an occurrence rate $r$ that is estimated from the total number of counts $n$ over the experiment duration.
The probability for $n$ events to occur within an observation window $\delta t$ is given by the Poisson distribution,
\begin{equation}\label{eq:pois}
\pois{r\delta t}{n} =  \frac{(r\delta t)^n e^{-r\delta t}}{n!}.
\end{equation}
We estimated qubit event and detector pulse rates based on observed counts throughout the experiment duration $T=N\dtcycle$, where $N$ is the number of observation windows of duration $\dtcycle$ each.
An individual count represents that there was at least one detection during the observation window.

Occurrence rates were calculated by considering that $n/N$ is an estimate for the probability to observe at least one event, or pulse, within an observation window (\mbox{$\pois{r \dtcycle}{n \ge 1}= 1 - e^{-r\dtcycle}$}),
\begin{equation}\label{eq:rate_approx1}
r = -\frac{1}{\dtcycle}\log\bigg(1 - \frac{n}{N}\bigg) \approx \frac{n}{T} 
\end{equation}
where the approximation requires \mbox{$(n/N)^2\ll 1$}, which is an applicable condition for all occurrence rates in the experiment.
Similarly, we calculated each expected number of counts for events, pulses, and their coincidences in the experiment as,
\begin{equation}\label{eq:rate_approx2}
n_\mathrm{expect} = N \cdot \pois{r \dtcycle}{n \ge 1}
\approx  \epsilon\sigma\Phi \cdot T
\end{equation}
where $r\dtcycle=\epsilon\sigma\Phi\dtcycle$ is expected number of cosmic-ray observations per observation window (Eq.~\ref{eq:rate}).

\subsubsection{Inter-arrival distribution background model}\label{sec:interarrival_bkgnd}
Since the occurrence rates for qubit relaxation events and detector pulses are described by Poisson random variables, 
we created an expected inter-arrival distribution for cosmic-ray coincidences, accidental (false-positive) coincidences, and other background inter-arrival delays outside the coincidence window. Accidental qubit-detector coincidences occur when a qubit event and a detector pulse are randomly within the same coincidence observation window (not caused by a common cosmogenic particle).
The probability of qubit-detector accidentals is calculated from the probability for two independent Poisson processes to each have at least one event within the same observation window of duration \mbox{$\delta t$}:
\begin{equation}\label{eq:bkgnd1}
P_\mathrm{QS}^\text{\tiny acc.}
=
\pois{r_\mathrm{Q}^\prime \delta t}{n \geq 1}
\pois{r_\mathrm{S}^\prime \delta t}{n \geq 1}
=
(1-e^{-r_\mathrm{Q}^\prime \delta t})
(1-e^{-r_\mathrm{S}^\prime \delta t}),
\end{equation}
where \mbox{$r_\mathrm{Q}^\prime=\rateQ - \rateCRQD$} and \mbox{$r_\mathrm{S}^\prime = \rateD - \rateCRQD\approx \rateD$} are the rates of qubit events and detector pulses, respectively, that are not caused by the same cosmogenic particle.
The rate of accidentals is then,
\begin{equation}\label{eq:bkgndQD0}
\bkgndQD 
\approx
\frac{1}{\delta t}P_\mathrm{QS}^\text{\tiny acc.}
\approx
(\rateQ - \rateCRQD) (1-e^{-\rateD \delta t}),
\end{equation}
where the approximations of \mbox{Equation~\ref{eq:rate_approx2}} are applicable since
\mbox{$(\bkgndQD \delta t)^2, (\rateQ\delta t)^2 \ll 1$}.

A calculation of the accidental rate from \mbox{Equation~\ref{eq:bkgndQD0}} includes the rate of  cosmic-ray coincidences $\rateCRQD$ which is calculated from the accidental rate itself (\mbox{$\rateCRQD=\rateQD - \bkgndQD$}).
We insert this relation into \mbox{Equation~\ref{eq:bkgndQD0}} and resolve the accidental rate:
\begin{equation}\label{eq:bkgndQD}
\bkgndQD = (\rateQ - \rateQD)(e^{\rateD \delta t}-1),
\end{equation}
which relies solely on rates that are calculated from observed counts~ (Eq.~\ref{eq:rate_approx1}).

We extend this background model for accidentals (Eq.~\ref{eq:bkgndQD}) to inter-arrival delay intervals outside the coincidence observation window.
The background inter-arrival delay distribution accounts for the likelihood that a qubit event and detector pulse occur within an observation window \mbox{$[\Delta t-\delta t/2,\Delta t+\delta t/2]$} for the inter-arrival delay \mbox{$\Delta t$}.
Since we condition inter-arrival durations on qubit events, we consider
the probability density for the nearest-in-time arrival of at least one detector pulse at the inter-arrival delay $t$ is,
\begin{equation}\label{eq:psd}
P_{\mathrm{S|Q}, t}^\text{\tiny bkgnd.}\,dt
=r_\mathrm{S}^\prime e^{-2r_\mathrm{S}^\prime |t|}\,dt
\approx\rateD e^{-2\rateD |t|}\,dt.
\end{equation}
We integrate Equation~\ref{eq:psd} over the observation window  $\delta t$ centered on the inter-arrival delay $\Delta t$,
\begin{equation}
P_{\mathrm{S|Q}, \Delta t}^\text{\tiny bkgnd.}
=\int_{\Delta t - \delta t / 2}^{\Delta t + \delta t / 2} \rateD e^{-2 \rateD |t|} dt = e^{-2 \rateD |\Delta t|} \sinh(\rateD\delta t),
\end{equation}
where we have evaluated for the condition that the inter-arrival delay is outside the coincidence window (\mbox{$|\Delta t| \ge \delta t/2$}).
From this probability we find the background rate for the inter-arrival delay interval similarly to Equation~\ref{eq:bkgnd1}:
\begin{equation}
r_{\mathrm{QS}, \Delta t}^\text{\tiny bkgnd.}
= (\rateQ - \rateCRQD) e^{-2 \rateD |\Delta t|} \sinh(\rateD\delta t),
\end{equation}
which is used to calculate the background distribution histograms displayed in Figure~\ref{fig:fig2} and Figure~\ref{fig:qb_det_coin}.

\subsubsection{Observation summary}\label{sec:observation_summary}
Table~\ref{tab:obs_det_qb} summarizes the observed and expected counts of \stc{} events, detector pulses, and their coincidences during the experiment duration.
The qubit array is denoted as \det{q} and each detector is listed according to its label~(Fig.~\ref{fig:fig1p0} and Table~\ref{tab:detectors}).
The label \det{s} represents ``any detector'' collectively
~\mbox{(\det{a} or \det{b} or ... or \det{f})},
the label \mbox{$d\text{-\det{s}}$} represents a coincidence of detector $d$ and any other detector~\mbox{(Eq.~\ref{eq:rate_any})}, 
and \det{ss} represents the coincidence of ``any two or more detectors''.

The observed expected counts were evaluated for an observation time window (Section~\ref{sec:rates}) of one measurement cycle for qubit events (\det{q}), detectors, and multi-detector coincidences.
Qubit-detector coincidence counts were evaluated for an observation time window of \mbox{$\dtcoin=3\,\mathrm{cycles}$} (Section~\ref{sec:qb_det_coin}).

For expected counts of the qubit array (\det{q}), we consider that every cosmic-ray impact to the qubit array is identifiable (i.e. \mbox{$\epsilon_\mathrm{Q}=1$} as in Eq.~\ref{eq:rateCRQD}).
For expected qubit-detector coincidences, we consider a coincidence identification efficiency of $\epCoin=\valueEffQD{}$ (Section~\ref{sec:qb_det_coin}) and account for the efficiencies of the scintillating detectors (Section~\ref{sec:eff_flux}).
The expected counts for qubit-detector coincidences are the sum of cosmic-ray and accidental contributions~\mbox{(Section~\ref{sec:interarrival_bkgnd})}.
\begin{table}[htbp]
\centering
\caption{\textbf{Observed and expected counts of qubit events, detector pulses, and their coincidences.}
All expected counts were calculated for the total duration of $\valueDurationHoursPrecise{}\,\mathrm{hours}$ and an integral muon flux \mbox{$\Phi=\valuePhi$}.
The fractional uncertainty of this flux is representative of the fractional uncertainty of the tabulated expected counts.
We note that expected and observed qubit event (\det{q}) counts differ significantly due to  events caused by non-cosmogenic sources.
}
\label{tab:obs_det_qb}
\begin{tabular}{rrr c rrr c rrrclc}
\hline
\multicolumn{1}{c}{} & \multicolumn{1}{c}{Observed} & \multicolumn{1}{c}{Expected}
& \hspace{0.25cm} & \multicolumn{1}{c}{} & \multicolumn{1}{c}{Observed} & \multicolumn{1}{c}{Expected}
& \hspace{0.25cm} & \multicolumn{1}{c}{} & \multicolumn{1}{c}{Observed} & \multicolumn{3}{c}{Expected} & \multicolumn{1}{c}{Accidentals} \\ \hline
A & 3,588,279 & 3,274,432 & & AS & 2,201,239 & 2,039,474 & & QA & 73 & 79.9 & $\pm$ & 4.8 & 1.6 \\
B & 3,515,419 & 3,616,208 & & BS & 2,488,386 & 2,667,358 & & QB & 98 & 94.1 & $\pm$ & 5.7 & 1.6 \\
C & 3,450,027 & 3,642,828 & & CS & 2,961,686 & 3,115,612 & & QC & 68 & 76.5 & $\pm$ & 4.6 & 1.6 \\
D & 3,492,167 & 3,655,762 & & DS & 2,376,675 & 2,438,638 & & QD & 57 & 57.6 & $\pm$ & 3.5 & 1.6 \\
E & 4,677,164 & 4,756,195 & & ES & 3,086,410 & 3,015,578 & & QE & 74 & 84.5 & $\pm$ & 5.1 & 2.1 \\
F & 4,663,288 & 4,399,031 & & FS & 2,492,331 & 2,398,536 & & QF & 71 & 63.8 & $\pm$ & 3.8 & 2.1 \\
S & 14,403,488 & 14,304,828 & & SS & 6,623,430 & 6,958,245 & & QS & 222 & 215.0 & $\pm$ & 12.4 & 6.4 \\
Q & {9,460} & 1,671 & &&& & &&&&
\end{tabular}
\end{table}

\subsection{Spatiotemporally Correlated Errors from a Manufactured Radiation Source}\label{sec:cs137}
We tested the expectation that the qubit array is sensitive to gamma radiation by exposing the qubit array to a source of ionizing gamma radiation (\Cs{}, \mbox{$17.2\,\mu\mathrm{Ci}$}, $662\,\mathrm{keV}$) placed outside the cryostat.

The \Cs{} source was positioned approximately \mbox{$20\,\mathrm{cm}$} from the qubit array along the $x$-axis~(Fig.~\ref{fig:fig1p0}).
We measured qubit relaxation for \mbox{$16.6\,\mathrm{hours}$} during \run{7} using the pulse sequence described in Section~\ref{sec:pulse_sequence} and detected \stc{} qubit relaxation events according to Section~\ref{sec:mf} (same data collection and analysis procedure as all other presented data).

In the presence of the \Cs{} source, we observed \stc{} qubit relaxation events at a rate of \mbox{$1/(10\pm 3\,\mathrm{s})$}.
The significant (10x) increase of \stc{} events compared to the baseline rate \mbox{$\rateQ=\valueRateQ{}$} (from all runs without an added source) shows that these events can result from gamma radiation impacts to the qubit array.
This is consistent with the claim that the \stc{} qubit relaxation events that are not accounted by cosmogenic sources are from other ionizing radiation sources in the laboratory and the experiment setup, such as gamma-ray emission from potassium or from nuclei of the uranium and thorium decay chains.

\subsection{Model of the Cosmic-ray Flux and Exposure Rate of Detectors}\label{sec:cr_model}
Our model of cosmic rays describes the occurrence rate of cosmogenic particles that are incident to the scintillating detectors and the qubit array.
The application of this model not only provides a comprehensive account of the cosmic-ray effect on our detectors, but also establishes a predictive tool for the rate of cosmic-ray  impacts to the qubit array.

Cosmogenic particles, such as muons, neutrons, protons, electrons, and neutrinos, are created in the upper atmosphere from primary cosmic rays (90\% protons, 9\% alpha particles, 1\% heavier nuclei)~\cite{Workman2022b}.
Our detection scheme is sensitive to \emph{simultaneous} energy depositions in the scintillators and qubit array substrate which are, overwhelmingly, caused by cosmogenic muons traversing the laboratory space (compared to protons, neutrons, pions, and other particles created by primary cosmic rays)~\cite{Workman2022b}.
In this work, we consider energy depositions from cosmogenic muons and the additional particles they create while traversing the laboratory space, such as knock-on electrons and gamma rays.

\subsubsection{Flux distribution and sampling of Cosmic-ray muons}\label{sec:cr_model_muon}
The cosmogenic muon flux distribution has a cosine-squared dependence on zenith angle $\theta$ and is independent of azimuthal angle $\varphi$.
We consider a differential flux distribution based on the energy distribution described by the Gaisser formula at zenith~\cite{gaisserCosmicRaysParticle2016, guan15},
\begin{equation}\label{eq:muon_distribution}
J_\mu (\theta, \varphi, E_\mu) = \frac{d\Phi}{dE_\mu d\Omega} = \frac{dI_\mu (\theta)}{dE_\mu} = C_\mu \bigg(\frac{E_\mu}{1\,\mathrm{GeV}}\bigg)^{-2.7}\bigg[\frac{1}{1+\frac{1.1E_\mu}{115\,\mathrm{GeV}}}+\frac{0.054}{1+\frac{1.1 E_\mu}{850\,\mathrm{GeV}}}\bigg] \cos^2\theta,
\end{equation}
where $\Phi$ is the total muon flux, $E_\mu$ is the muon energy (10~GeV to 1~TeV), $\Omega$ denotes solid angle, $I_\mu$ is the muon flux integrated over energy, and $C_\mu$ is a free scale factor that is absorbed into the overall muon flux $\Phi$ upon model calibration (Section~\ref{sec:det_cal}).
Although the Gaisser formula is most applicable for muon energies \mbox{$\gtrsim (100\,\mathrm{GeV}/\cos\theta$)} and small zenith angles, this model is a reasonable approximation for our experiment.
Since these high-energy muons are minimum ionizing, the energy they deposit in a detector or the qubit array substrate depends on the distance traveled through it~\cite{Workman2022b}.
Consequently, our results are rather insensitive to the muon energy distribution associated with the angular distribution of our model~(Eq.~\ref{eq:muon_distribution}).
We additionally compared the zenith-dependent flux of our model \mbox{(Eq.~\ref{eq:muon_distribution})} to another distribution that accounts for the Earth's curvature and the muon energy-dependence on zenith angle~\cite{guan15}.
We found negligible differences relevant to energy deposition rates to detectors and the qubit array substrate between these models, especially for the zenith angles related to the rate of coincidence events in our experiment.

\begin{figure}[htbp]
\includegraphics{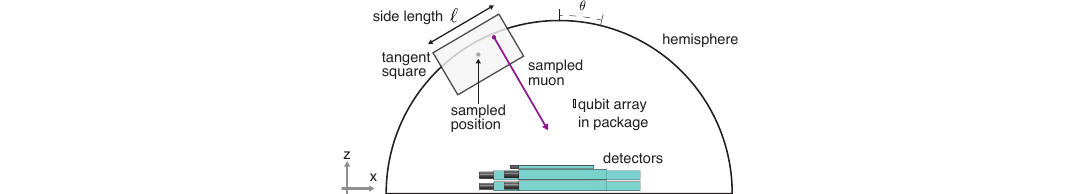}
\caption{\label{fig:fig_hemisphere} \textbf{Illustration of the cosmogenic muon sampling technique.} For each randomly sampled muon we sampled the muon energy, the zenith and azimuthal angle of a tangent square on the hemisphere (not to scale), and the muon position within the tangent square.
The muon is thrown toward the detectors with a momentum vector that is parallel to the tangent square normal vector.
}
\end{figure}

We performed Monte Carlo simulations of cosmic rays in our laboratory using the \geant{} toolkit (Section~\ref{sec:g4}).
We sampled individual cosmic-ray muons from the flux distribution of \mbox{Equation~\ref{eq:muon_distribution}} to estimate the exposure of the detectors to cosmic-ray muons.
(The qubit array is also considered a detector in this regard.)
This sampling technique ultimately provided cross-sections for the rate of energy depositions into each detector, the qubit array, and their coincidences 
within a specific energy range.

We used the following sampling procedure to calculate the interaction cross-sections.
We define a hemisphere with a radius (\mbox{$15\,\mathrm{m}$}) that surrounds the scintillating detectors and qubit array (Fig.~\ref{fig:fig_hemisphere}) along with other objects in the laboratory environment~\mbox{(Section~\ref{sec:g4})}.
The procedure for sampling each muon is
\begin{enumerate}
    \item Randomly sample a position on the hemisphere by choosing zenith and azimuthal angles from \mbox{$(\cos^2 \theta \sin \theta)$} and uniform \mbox{$\varphi\in [0,2\pi)$} distributions, respectively.
    \item Define a square of side-length $\ell$ that is centered on, and tangent to, the sampled position. 
    (The side-length $\ell$ is a fixed length for all samples. Ensure that \mbox{$\ell > 2d$} where $d$ is the farthest distance of the detector objects from the hemisphere origin that is relevant for particle interactions.)
    \item Randomly sample a point within the extent of the tangent square.
    \item Randomly choose a muon energy according to the muon flux distribution~\mbox{(Eq.~\ref{eq:muon_distribution})}.
    \item ``Throw'' the muon from the sampled point in the direction normal to the tangent square toward the detectors.
    \item Propagate the muon trajectory and its interactions within the laboratory~(Section~\ref{sec:g4}).
    \item Record the total energy deposited in each detector.
\end{enumerate}

\subsubsection{Energy-deposition occurrence rate and interaction cross-sections}\label{sec:cross_section}
Interaction cross-sections $\sigma$ relate an occurrence rate $r$ of energy depositions and the total cosmogenic muon flux $\Phi$ as
\begin{equation}\label{eq:rate}
r = \epsilon \sigma \Phi,
\end{equation}
where $\epsilon$ represents the efficiency of the detector(s).
For a detector, or combination of detectors, the cross-section $\sigma$ is calculated from the sampled energy depositions (Section~\ref{sec:cr_model_muon}), 
\begin{equation}\label{eq:beta}
\sigma = \frac{\text{\# of energy depositions to the detector(s)}}{\text{\# of muons sampled}/\ell^2}.
\end{equation}
Each cross-section has a geometric quality, as it depends on the dimensions of the detector(s) and, in the case of coincidence combinations, the relative placement of detectors.

Coincidence events result from particle energy depositions to multiple detectors from one initially sampled muon.
Similar to Eq.~\ref{eq:rate}, the occurrence rate for multi-detector coincidences is
\begin{equation}\label{eq:rate_coin}
r=\bigg(\prod_{k\in\alpha} \epsilon_k \bigg)\sigma\Phi,
\end{equation}
where the efficiency factor is the product of the efficiencies for each detector $k$ within the combination $\alpha$.
The cross-section for a multi-detector coincidence is calculated according Equation~\ref{eq:beta} by counting the number of sampled muons that each resulted in an energy deposition to every detector within the coincidence combination.
For example, the cross-section for a coincidence of detector \det{a} and detector \det{b} is,
\begin{equation}\label{eq:beta_inclusive_num}
\sigma_\mathrm{AB} = \frac{\text{\# of energy depositions to both \det{a} and \det{b}}}{\text{\# of muons sampled}/\ell^2},
\end{equation}
where the numerator counts the number of sampled muons that resulted in a coincidence of both detectors \det{a} and \det{b}.
For the experimental analysis, we calculated all cross-sections for a specific range of energy depositions for each detector, matching the acceptance criteria for pulse detection (Table~\ref{tab:detspec}).

We refer the cross-section $\sigma$ as an \textit{inclusive} cross-section (as defined in Eqs.~\ref{eq:beta}~and~\ref{eq:beta_inclusive_num}) since it corresponds to a detector combination that is mutually inclusive with other detector combinations that could also occur from the same muons.
We define \textit{exclusive} cross-sections $\sigma^*$, that correspond to the set of coincidence observations
that are mutually exclusive and completely exhaustive among all detectors included in the simulation.
An exclusive observation for a given combination requires there are no energy depositions to the other detectors.
For example, the exclusive cross-section for the coincidence of detectors \det{a} and \det{b} requires that the energy depositions to \det{a} and \det{b} occur when there are no energy depositions to other detectors (\mbox{\{\det{c}, \det{d}, \det{e}, \det{f}, \det{q}\}}):
\begin{equation}\label{eq:beta_exclusive_num}
\sigma_\mathrm{AB}^* = \frac{\text{
\# of energy depositions to \det{a} and \det{b} but not \det{c}, \det{d}, \det{e}, \det{f}, nor \det{q}
}}{\text{\# of muons sampled}/\ell^2}.
\end{equation}
While an inclusive combination $\alpha$ only relies on the detectors within that combination, an exclusive combination $\alpha$ depends on every detector in the analysis.
The set of exclusive cross-sections have the property \mbox{$\sum_\alpha\sigma_\alpha^*=\ell^2$,}
where $\alpha$ sums over all combinations of the detectors (including the null cross-section for no energy deposition to any detector).

The exclusive and inclusive cross-sections are related by linear combinations.
For example, the inclusive cross-section for the combination $\alpha$ is the sum of 
the exclusive cross-section $\sigma_\alpha^*$ and all other exclusive combinations that include $\alpha$ with other detectors:
\begin{equation}
\sigma_\alpha = \sigma_\alpha^* + \sum_{\alpha^\prime}\sigma_{\alpha^\prime}^*,
\end{equation}
where the sum is over all combinations $\alpha^\prime$ such that $\alpha\in\alpha^\prime$ (meaning that $\alpha^\prime$ contains all the detectors in $\alpha$).

In the context of modeling experiment observations, our estimated value of a cross-section can have error.
Inclusive cross-section errors are positively correlated, e.g.~\mbox{$\delta\sigma_\mathrm{A}\propto \delta\sigma_\mathrm{AB}$}, which states that an over-estimate of the \det{ab} coincidence cross-section connotes an over-estimate of the \det{a} cross-section.
In contrast, exclusive cross-sections can have positive or negative correlations, \mbox{e.g.~$-\delta\sigma_\mathrm{A}^*\propto \delta\sigma_\mathrm{AB}^*$}, which states that an over-estimate of the exclusive \det{ab} coincidences contributes to an under-estimate of the \det{a} exclusive cross-sections. 
In general, systematic error of any particular exclusive cross-section affects the error of all other exclusive cross-sections, resulting in both over- and under-estimated rates for any particular exclusive combinations we may observe.

We consider coincidence observations that are generally insensitive to the error of any particular cross-section by summing over exclusive cross-section combinations.
Specifically, we refer to the rate of (two or more)-fold coincidences of detector $d$ and any other \mbox{detector \det{s}:}
\begin{equation}\label{eq:rate_any}
r_{d\text{-}\mathrm{S}}=
\epsilon_d\sum_{\alpha}\bigg(1 - \prod_{\substack{k\in\alpha\\ k\neq d}}\overline{\epsilon}_k \bigg)\sigma_{\alpha}^*\Phi,
\end{equation}
where the sum is over all combinations $\alpha$ that contain detector $d$ with other detectors, each product is over detectors \mbox{$k\in\alpha$} (except $d$), and $\overline{\epsilon}_k=1-\epsilon_k$ are detector inefficiencies.
The products within Equation~\ref{eq:rate_any} account for the rate that at least one detector of the combination $\alpha$ (except $d$) produces a pulse provided there are coincident energy depositions to the detectors in the exclusive combination $\alpha$.
We define the coverage of detector $d$ provided by all other detectors,
\begin{equation}\label{eq:coverage_d}
C_\ds=\frac{r_\ds}{r_d}
=\frac{1}{\sigma_d}
\sum_{\alpha}\bigg(1 - \prod_{\substack{k\in\alpha\\ k\neq d}}\overline{\epsilon}_k \bigg)\sigma_{\alpha}^*,
\end{equation}
which again includes the sum over all combinations $\alpha$ such that \mbox{$d\in\alpha$} and each product is taken over detectors \mbox{$k\in\alpha$ (except $d$)}.
We remark that the coverage $C_\ds$ is independent of both the muon flux and the efficiency of detector $d$.

\subsection{GEANT4 Simulation}\label{sec:g4}
\geant{} can simulate particle transport and interactions with matter for applications in high-energy and nuclear physics~\cite{alli16}.
We used \geant{} to propagate cosmogenic muon trajectories, simulate their interactions with objects in the laboratory, and predict cross-sections for energy depositions to the detectors and qubit array~\mbox{(Section~\ref{sec:cross_section})}.
We calibrated each detector's energy response based on the predicted distributions of deposited energy produced by the \geant{} simulations.
The calibration of each detector enabled an accurate calculation of all interaction cross-sections among the detectors and the qubit array.

\begin{figure}[htbp]
\includegraphics[width=1.0\textwidth]{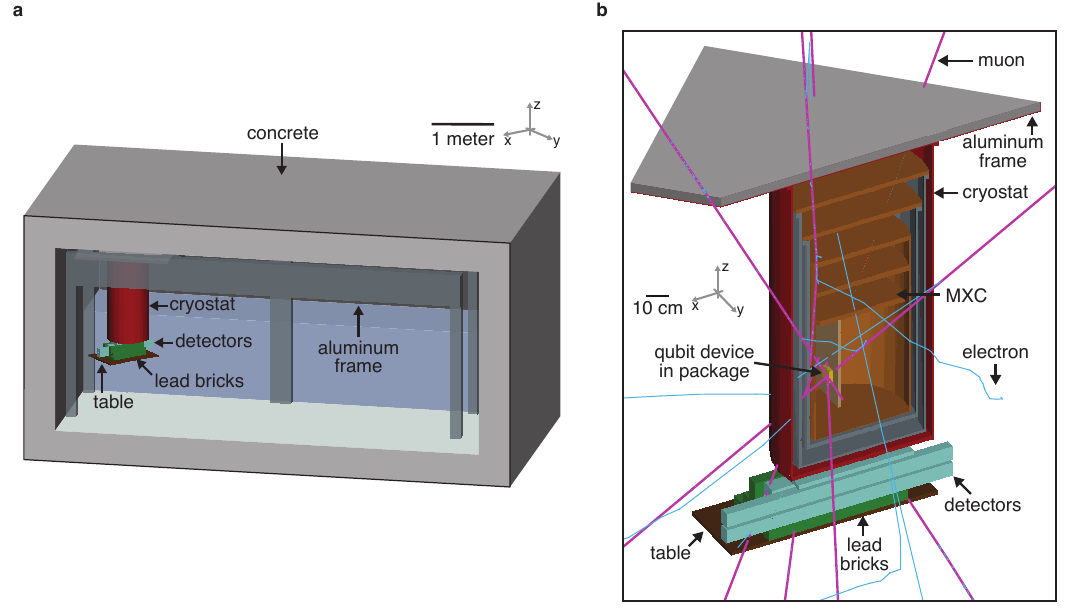}
\caption{\label{fig:g4_scene}
\textbf{Model of the laboratory and experiment setup simulated in GEANT4.}
(a) The physical model that was simulated in \geant{} included concrete surrounding the laboratory and objects near the cryostat and detectors that could influence energy depositions.
(b) The particle tracks created by five cosmogenic muons (purple) are shown propagating through the cryostat (cross-sectional view).
\geant{} registered if each particle resulted in an energy deposition to detectors (cyan) or the qubit array substrate located inside its package (gold).
The cosmic-ray muons originate from outside the laboratory scene and create secondary particles (electrons, blue) from interactions with objects in the laboratory, including 
the aluminum frame (gray), 
vacuum can (red),
the cryostat stages and mixing chamber paddle (copper), lead bricks (green), and a wooden table (brown).
}
\end{figure}

Cosmogenic muons that traverse through dense objects create additional particles~\cite{Workman2022b} that contribute to energy depositions into the detectors and qubit array substrate.
We used \geant{} to model these interactions of high-energy muons with the detectors and qubit array substrate, as well as laboratory infrastructure (Fig.~\ref{fig:g4_scene}a). These objects included
the concrete ceiling, walls, and floor of the laboratory (\mbox{$0.5\,\mathrm{m}$} thickness),
the aluminum frame that supports the cryostat,
the experiment cryostat with internal structure,
the qubit array package,
and lead bricks near the detectors.
Since high-energy muons interact electromagnetically with these objects~\cite{Workman2022b},
we used the pre-selected QBBC physics list of \geant{} with the EMV option. % https://geant4-userdoc.web.cern.ch/UsersGuides/PhysicsListGuide/html/index.html
This physics list is relevant for the charged particle transport associated with cosmogenic muon scattering and creation of the secondary particles that they create (knock-on electrons, gamma rays, etc.).

We performed the simulation by sampling the energies and directions (Eq.~\ref{eq:muon_distribution}) of positively charged cosmogenic muons on a hemisphere surrounding the laboratory space, as described by the procedure in Section~\ref{sec:cr_model_muon}.
Figure~\ref{fig:g4_scene}b shows the particle tracks from five independently sampled cosmogenic muons (purple lines).
The muon interactions with the laboratory infrastructure resulted in the production of knock-on electrons (blue lines).
Gamma rays are also produced from local particle interactions (omitted from the illustration for clarity).
Our overall simulation results show non-muon particles contributing to {4-9\%} of energy depositions among the scintillating detectors (for the accepted energy range) and \valueSigmaOtherFractionQ{} of all cosmogenic energy depositions to the qubit array.

For every cosmogenic muon that was thrown at the experiment setup, the \geant{} simulation informed if, and how much, energy was deposited into each detector and the qubit array substrate.
A small subset of the simulation results are shown in Table~\ref{tab:g4_sim_dataframe} to convey how \geant{} results were organized.
Each row of the table represents the energy depositions from one muon (of $N$) that was sampled.
Each detector (and qubit array) is assigned to a column.
For example, the cosmic-ray muon sampled for $n=3$ in this table resulted in a four-fold coincidence of energy depositions into detectors \det{b}, \det{c}, \det{d}, and the qubit array \det{q}.

We categorized the energy depositions within each column by the type of particle that initially entered the detector volume (even if additional particles are created inside the detector).
The particle assignments were classified as ``$\mu$'' (indicative of a muon) or ``other'' (for any non-muon particle).
These specific categories were created to differentiate interaction rates originating from the initially sampled muon and those from secondary particles produced in the laboratory environment.

Interaction cross-sections were calculated from the number of initially sampled muons that resulted in at least one energy deposition to each detector in a coincidence combination (Eq.~\ref{eq:beta}).
The total deposited energy per sampled muon, summed over both muon and non-muon particles, was used when calculating cross-sections for a specified energy range (Section~\ref{sec:det_cal}). 
The total energy deposited into a given detector (from each sampled muon) is categorized as ``muon'' if any one muon particle enters the detector (e.g.~the energy deposition combinations in each row of Table~\ref{tab:g4_sim_dataframe} are all categorized as muon energy depositions).

\begin{table}[htbp]
\centering
\caption{\label{tab:g4_sim_dataframe}
\textbf{Example simulation results from GEANT4.} Energy depositions to the detectors and qubit array (columns) substrate were recorded for sampled cosmic-ray muons (rows).
All energies are in units of MeV. The symbol ``--'' denotes ``no energy deposited''. }
\begin{tabular}{c|cc|cc|cc|cc|cc|cc|cc}
\hline
\multirow{2}{*}{$n$} & \multicolumn{2}{c|}{Det. A} & \multicolumn{2}{c|}{Det. B} & \multicolumn{2}{c|}{Det. C} & \multicolumn{2}{c|}{Det. D} & \multicolumn{2}{c|}{Det. E} & \multicolumn{2}{c|}{Det. F} & \multicolumn{2}{c}{Det. Q} \\ 
 & $E_{\mu}$ & $E_{\mathrm{other}}$ & $E_{\mu}$ & $E_{\mathrm{other}}$ & $E_{\mu}$ & $E_{\mathrm{other}}$ & $E_{\mu}$ & $E_{\mathrm{other}}$& $E_{\mu}$ & $E_{\mathrm{other}}$& $E_{\mu}$ & $E_{\mathrm{other}}$& $E_{\mu}$ & $E_{\mathrm{other}}$\\ \hline
0	&   --   &   --   &   7.33   &   --   &   5.00   &   --   &   --   &   --   &   --   &   --   &   --   &   --   &   --   &   --   \\
1	&   3.16   &   --   &   --   &   --   &   --   &   --   &   --   &   --   &   15.50   &   3.43   &   0.03   &   4.77   &   --   &   --   \\
2	&   4.36   &   --   &   --   &   --   &   --   &   --   &   --   &   --   &   --   &   --   &   --   &   --   &   --   &   --   \\
3	&   --   &   --   &   4.00   &   --   &   14.05   &   --   &   11.9   &   --   &   --   &   --   &   --   &   --   &   0.24   &   --   \\
4	&   --   &   --   &   --   &   --   &   17.08   &   --   &   --   &   --   &   --   &   --   &   --   &   --   &   --   &   --   \\
\vdots & \multicolumn{14}{c}{\vdots} \\
\end{tabular}
\end{table}

We used the methods described above to perform two simulations. 
First, we performed a \geant{} simulation for energy depositions among the detectors alone.
For this detector-focused simulation, we sampled \mbox{$N=30\cdot10^6$} muons from a tangent square of side-length \mbox{$\ell=2\,\mathrm{m}$}, and starting from a random position on the hemisphere (primary vertex at coordinate \mbox{$(0, 0, -0.6)\,\mathrm{m}$}, with the qubit chip positioned at the origin~\mbox{(Section~\ref{sec:cr_model_muon})}.
Since the qubit array is small relative to the detectors, the tangent square area was too large for sufficient sampling of qubit-detector coincidences. 
The results from this simulation were used for calibration of the detectors and estimation of the muon flux in the laboratory (Section~\ref{sec:det_cal}).

We performed an additional \geant{} simulation for energy depositions among the qubit array and detectors.
For this qubit array-focused simulation, we sampled \mbox{$N=40\cdot10^6$} muons and used importance sampling by centering the hemisphere's primary vertex in the qubit array substrate and by decreasing the tangent square side-length to \mbox{$\ell=10\,\mathrm{cm}$}.
Although this importance sampling technique biases the energy deposition rates among the detectors, it does not bias the rate of qubit-detector coincidences. 
We evaluated the qubit array cross-section and all qubit-detector coincidence cross-sections based on the results of this qubit array-focused simulation.

\subsection{Detector Calibration}\label{sec:det_cal}
We calibrated the energy response of each scintillating detector
by relating measured distributions of detector pulse amplitude with expected distributions of deposited energy from \geant{}.
This enabled an accurate calculation of all interaction cross-sections, including the cross-sections for  qubit-detector coincidences.
We then estimated the detection efficiency from the coverage that the detectors provide each other.
The muon flux was estimated from the measured rate of (two or more)-fold coincidences.

\subsubsection{Detector energy-response model}
Once radiation ionizes atoms in the scintillating polymer of a detector, a flash of indigo to blue light (420-450 nm) is produced with an intensity proportional to the energy deposited $E$ (MeV units) to the detector.
The light is transformed into an electrical pulse by a photomultiplier tube (PMT) linked to the scintillator. 
After amplification, an FPGA-based analog-to-digital converter discriminates the pulse, performs a trapezoid filter, then records the pulse amplitude $\Eadc$ (arbitrary units).
For each detector, we assert a proportional relationship between the pulse amplitude $\Eadc{}$ and an estimated energy $\EMeV$ (MeV units) referred to the voltage~\cite{yane21}:
\begin{equation}\label{eq:det_cal_model_1}
\Eadc{} = \VtoE{} \EMeV,
\end{equation}
where $\VtoE{}$ is the proportionality factor between the voltage-referred energy and the pulse amplitude.

The voltage-referred energy $\EMeV{}$ differs from the energy deposited $E$ since each detector has imperfect energy resolution.
A distribution of voltage-referred energy $P(\EMeV)$ is related to its corresponding distribution of energy deposited $P(E)$ by a Gaussian convolution:
\begin{equation}\label{eq:det_cal_model_2}
P(\EMeV{})\,d\EMeV{}=
\bigg(\int_{E}dE\,P(E) \frac{1}{\sigma(E)\sqrt{2\pi}} e^{-\frac{(E - \EMeV)^2}{2\sigma(E)^2}}
\bigg)\,d\EMeV.
\end{equation}
The convolution includes an energy-dependent resolution defined as~\cite{knol00}
\begin{equation}\label{eq:det_cal_model_3}
\sigma(E) = \Eres\sqrt{E_0 E},
\end{equation}
where $\Eres{}$ is the fractional energy resolution of the detector at the chosen reference energy \mbox{$E_0 = 5\,\mathrm{MeV}$}.
The pulse amplitude distribution $P(\Eadc{})$ is calculated from Equations~\ref{eq:det_cal_model_1}-\ref{eq:det_cal_model_3}, provided $\VtoE$, $\Eres$, and the expected distribution of energy deposited $P(E)$.

\begin{figure}[htbp]
\includegraphics[width=1.0\textwidth]{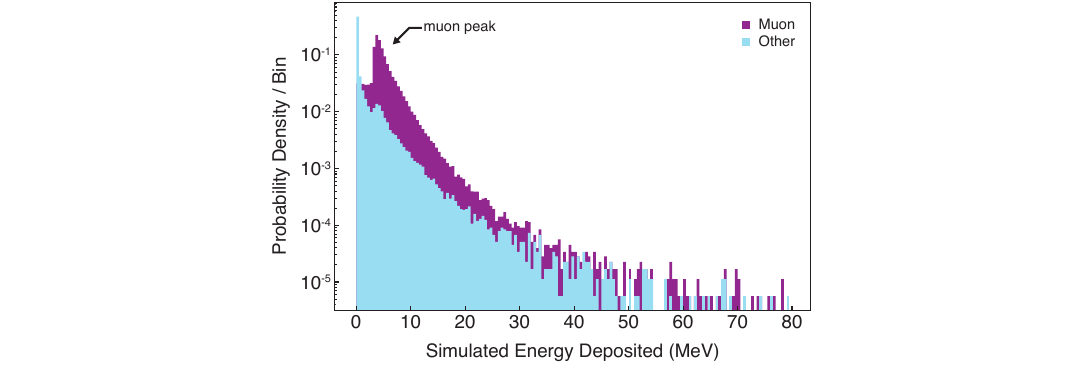}
\caption{\label{fig:fig_det_sim_spec} \textbf{Example energy spectrum from simulation.}
The expected PDF of energy deposited $P_{\mathrm{A}}(E)$ in detector \det{a}.
The \geant{} simulation results for detector \det{a} were binned, interpolated, and normalized.
Here, the PDF is shown before interpolation and is separated into muon and other particle energy depositions to display the relative contribution of these particles.
}
\end{figure}

From the \geant{} simulation results (Section~\ref{sec:g4}), we created expected probability distribution functions (PDFs) $P_{\alpha,k}(E)$ of the energy deposited to each detector $k$ given the coincidence combination $\alpha$ shared among the scintillators.
Each PDF $P_{\alpha,k}(E)$ was generated from the interpolation and the normalization of binned counts of energy deposited to a given detector.
Figure~\ref{fig:fig_det_sim_spec} shows an example PDF for the energy deposited to detector \det{a} calculated from \geant{} simulation results.
Since cosmogenic muons are minimum ionizing, the muon peak feature corresponds to the most probable straight line distance of a muon traversing through the detector.
The PDF for energy depositions from other particles tend toward lower energies below the muon peak as these particle energy depositions result from bremsstrahlung radiation of electrons and include low energy gamma rays (\mbox{$<1\,\mathrm{MeV}$}).

The expected pulse rate per pulse amplitude is calculated from the pulse amplitude PDF $P_{\alpha,k}(\Eadc)$ and the pulse rate $r_\alpha$ of the combination $\alpha$, given the efficiencies $\epsilon_j$ for the detectors in the combination $\alpha$, the cross-section $\sigma_\alpha$ for the combination, and the muon flux $\Phi$:
\begin{equation}\label{eq:cal_scale}
r_{\alpha,k}(\Eadc)\,d\Eadc=
r_\alpha \cdot P_{\alpha,k}(\Eadc)\,d\Eadc
=\bigg(\prod_{j\in\alpha}\epsilon_{j}\bigg)\sigma_\alpha \Phi
\cdot P_{\alpha,k}(\Eadc)\,d\Eadc,
\end{equation}
Equation~\ref{eq:cal_scale} is converted into expected counts, $N_{\alpha,k}(V_\mathrm{ADC}^i)$, for an $\Eadc$ bin $i$ by multiplying by the total duration (using Eq.~\ref{eq:rate_approx2}, since each observation window $\delta t$ is sufficiently small) and integrating over the pulse amplitudes of the bin.

\subsubsection{Detector model calibration}\label{sec:det_cal_subsec}
We performed a fitting procedure to estimate energy-response model parameters $\VtoE_k$ and $\Eres_k$ for each of the six detectors.
These parameters were estimated by iterative least squares minimization of a cost function
evaluated from expected (Eqs.~\ref{eq:det_cal_model_1}-\ref{eq:cal_scale}) and measured pulse amplitudes for the first data run (\run{0}, containing \mbox{$19\,\mathrm{hours}$} of detector pulse data total, collected asynchronously with qubit measurements).
The fit optimization procedure finds parameters $\VtoE_k$ and $\Eres_k$ that effectively stretch and broadens the PDFs from \geant{} to match the data.

We performed this fitting procedure by leveraging inter-dependencies of coincidence observations, which are determined by the dimensions and relative placement of the detectors.
The detector energy-response parameters, $\VtoE_k$ and $\Eres_k$, are constrained by the shape of the pulse amplitude distributions which generally depend on the direction muons traverse the detectors for each coincidence combination.
For the fitting procedure, we accounted for the inter-dependency of coincidence rates (Eq.~\ref{eq:cal_scale}),
by evaluating interaction cross-sections for all energy depositions simulated in \geant{} (without any constraint on deposited energy).

We estimated the 12 parameters, $\VtoE_k$ and $\Eres_k$ by simultaneous fitting \mbox{$\mathcal{O}(10^2)$} amplitude bins within each pulse amplitude distribution.
The fitted data includes the six pulse amplitude distributions from the individual detectors and 18 pulse amplitude distributions from each detector participating in eight coincidence combinations.
We chose the combinations that occurred within each detector stack ($\det{bc}, \det{cd}, \det{ad}, \det{bcd},\det{ae}, \det{ef}, \det{af}, \det{aef}$, and the six individual detectors) since these combinations had relatively significant occurrence rates and sufficiently constrain the energy-response model.

We included detection efficiencies $\epsilon_k$ and the muon flux $\Phi$ of Eq.~\ref{eq:cal_scale} as additional free parameters in the fitting procedure.
The efficiency and flux produced from the detector energy-response calibration have estimation error that resulted from a mismatch of the data and model for specific two-fold and three-fold coincidence combinations.
In the context of the energy-response calibration, the efficiencies and flux are nuisance parameters that are not strongly correlated with the $\VtoE_k$ and $\Eres_k$ parameter estimates for amplitude-to-energy conversion and detector resolutions, respectively.
After the detector calibration, we calculated cross-sections based on the fitted amplitude-to-energy factor and detector resolution and then evaluated the efficiencies and muon flux for the accepted energy range, considering the coincidence observables relevant in our experiment (Section~\ref{sec:eff_flux}).

The parameters were estimated by minimizing the cost function,
\begin{equation}\label{eq:cost}
-\log(\mathcal{L})=
\sum_{\alpha}
\sum_{k\in\alpha}
\sum_{i\in k}
\left[\mathcal{E}_{\alpha,k}^i - \mathcal{O}_{\alpha,k}^i + \mathcal{O}_{\alpha,k}^i \log (\mathcal{O}_{\alpha,k}^i) - \mathcal{O}_{\alpha,k}^i \log (\mathcal{E}_{\alpha,k}^i) \right],
\end{equation}
where $\mathcal{E}_{\alpha,k}^i$ and $\mathcal{O}_{\alpha,k}^i$ are the expected and observed counts, respectively, for the pulse amplitude bin $V_\mathrm{ADC}^i$ of each detector $k$ in the coincidence combination $\alpha$.
This cost function (Eq.~\ref{eq:cost}) is the sum of negative log-likelihood functions for all fitted bins.
Each function is the negative log ratio of likelihood functions for Poisson distributed random variables $\mathcal{E}_{\alpha,k}^i$ and $\mathcal{O}_{\alpha,k}^i$.
Since $-2\log(\mathcal{L})$
is distributed as $\chi^2$ if the counts in each bin is sufficiently large, this construction is advantageous for estimating uncertainty of the parameters \mbox{$\VtoE_k$ and $\Eres_k$}.

We evaluated the cost function over a chosen range of pulse amplitudes for each detector which are also the accepted pulse amplitudes for the qubit-detector coincidence analysis (Table~\ref{tab:detspec}).
The lower amplitude threshold rejects energy depositions from terrestrial radiation, while the upper amplitude threshold rejects pulses that are nonlinear in energy (Eq.~\ref{eq:det_cal_model_1}) due to charge pre-amplifier saturation.

Before fitting data, we identified detector coincidences within an inter-arrival window of $1\,\mu\mathrm{s}$ conditioned on the pulses of each detector, which has a negligible contribution of accidental coincidences.
Likewise, we identified detector coincidences in the \geant{} simulation results.
The fitted range of pulse amplitudes had an unknown relation to its corresponding range of deposited energy before the detector response parameters, \mbox{$\VtoE_k$ and $\Eres_k$}, were estimated by the fit optimization.
For this reason, we did not apply any filtering to the measured pulse amplitudes or simulated deposited energy before coincidence event identification.

Accordingly, we calculated expected counts (Eq.~\ref{eq:cal_scale}) using cross-sections for the entire range of deposited energy (evaluated  with Eq.~\ref{eq:rate} from sampled energy depositions).
Although the cost function (Eq.~\ref{eq:cost}) was evaluated for a prescribed amplitude range,
coincidence data within the range has correlation with measured pulse amplitudes (also expected energy deposited) outside the fitted range.
This implies that goodness-of-fit (for amplitude-to-energy conversion) partially relied on the agreement of observed and expected coincidences rates from data outside the fitted range of amplitudes.
\begin{figure}[htbp]
\includegraphics{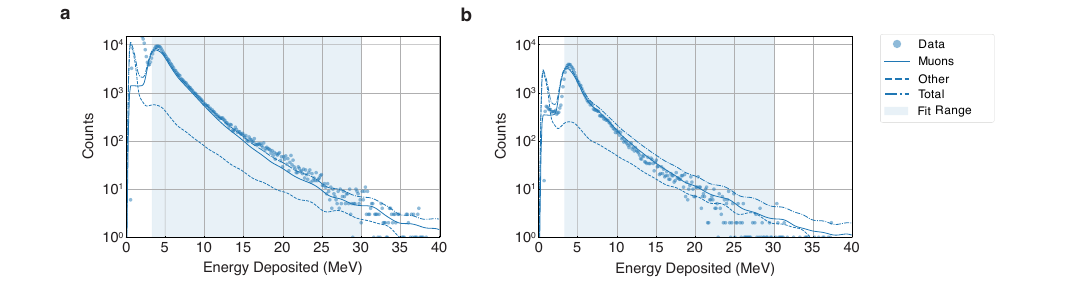}
\caption{\label{fig:fig_det_cal}
\textbf{Measured and expected distributions of deposited energy.} 
(a) The measured and expected energy distributions of detector \det{a}. The lowest energies show a relatively high rate of detector pulses from gamma radiation.
(b) The measured and expected energy spectra of detector \det{a} conditioned on \det{aef} coincidences.
}
\end{figure}

Figure~\ref{fig:fig_det_cal} shows a subset of measured and expected deposited energy distributions of detector \det{a} that were calculated after the fitting procedure.
The measured and expected distributions closely agree within the fitted range (shaded blue).
The modeled distributions align with the measured shape and energy of the muon peaks and the distribution's tail at higher energies.

We also consider the agreement between the data and model \textit{outside} the fitted energy region.
At the lowest energies outside the fitted region, our model does not describe data of a single detector alone (Fig.~\ref{fig:fig_det_cal}a) which is not conditioned on a coincidence.
Most of this anticipated discrepancy is from the flux of low-energy non-cosmogenic radiation which is primarily from gamma-ray radiation in the laboratory.
As shown in the coincidence distribution of Figure~\ref{fig:fig_det_cal}b, the \geant{} model predicts low-energy peaks from knock-on electrons produced by cosmogenic muons.
For deposited energies above the muon peak, the fitted model tends to overestimate the rate of \det{aef} coincidences relative to the measured counts, although this deviation may result of the nonlinear response of detector \det{a}.
The fraction of secondary particles predicted by \geant{} (labeled as ``other'', dotted line) is approximately $\lesssim 10\%$ within the fit region and representative of other detectors.

We note that detector pairs (\det{a} and \det{b}, \det{c} and \det{d}) have different energy resolution even though detectors within each pair have a similar construction (Section~\ref{sec:detectors}).
These differences may result from the sensitivity of the fitted energy resolution to the full-width half-maximum of the muon peak for a given detector, which differs among detectors.
Nevertheless, our overall experiment findings (related to qubit-detector coincidences) are generally insensitive to the absolute accuracy of the energy resolution parameters $b_k$, since accurate cross-section calculations are weakly affected by energy resolution.
The proportionality factors $a_k$ for energy-to-amplitude conversion
determine cross-section accuracy, as this affects the number of simulated energy depositions associated to the accepted range of pulse amplitudes.
\begin{table}[htbp]
\centering
\caption{\textbf{Detector model parameters.} 
The detector response parameters fitted to \run{0} were found to be consistent across all runs.
The fitting procedure yielded detector efficiency nuisance parameters in the range 88-95\%.
Uncertainty represents the standard error as reported by the least squares minimizer.
}
\label{tab:detspec}
\begin{tabular}{c  r@{$\,\pm\,$}l  r@{$\,\pm\,$}l  r@{$,\,$}l  r@{$,\,$}l}
\hline
\multirow{2}{*}{Detector}	& \multicolumn{2}{c}{Energy to $\Eadc$} & \multicolumn{2}{c}{Energy Resolution} & \multicolumn{4}{c}{Energy Range Limits} \\
							& \multicolumn{2}{c}{$\VtoE{}$ (\adc{}/MeV)} & \multicolumn{2}{c}{@ 5 MeV, $\Eres{}$ (\%)} & \multicolumn{2}{c}{$\Eadc$ (a.u.)} & \multicolumn{2}{c}{$\EMeV$ (MeV)}\\ \hline
A & \hspace{0.7cm} 14.971 & 0.026\hspace{1cm} & 6.3 & 0.4 &\hspace{0.4cm} 50 & 450 &\hspace{0.2cm} 3.4 & 30.4\\
B & 12.219 & 0.008& 1.5 & 0.04 		 & 35 & 400 & 2.9 & 32.7\\
C & 14.219 & 0.009& 1.6 	& 1.0 		 & 170 & 550 & 12.0 & 38.7\\
D & 14.832 & 0.011& 11.3 	& 0.3 		 & 170 & 550 & 11.5 & 37.1\\
E & 17.465 & 0.012& 11.8 	& 0.2 		 & 170 & 400 & 9.8 & 23.0\\
F & 21.700 & 0.014& 8.4 	& 0.2	 	 & 200 & 400 & 9.2 & 18.4\\ \hline
\end{tabular}
\end{table}

We evaluated cross-sections for the qubit-detector coincidence analysis after the detector energy-response calibration.
This procedure is consistent with the qubit-detector coincidence identification procedure, for which we first filtered for detector pulses within the accepted energy range \mbox{(Table~\ref{tab:detspec})} and then identified coincidences. 

\subsubsection{Estimation of detector efficiencies and muon flux}\label{sec:eff_flux}
We estimated the detector efficiencies from observed multi-detector coincidence rates and the total pulse rate of each detector.
For these calculations we used all detector data that was synchronously measured with qubit measurements (Section~\ref{sec:ref_timing}).

For each detector $d$, we evaluated the observed coverage \mbox{$C_\ds = r_\ds/r_d$} provided by all other detectors based on observed counts, where 
$r_d$ is the measured rate of pulses from detector $d$ and
$r_\ds$ is the measured rate of (two or more)-fold coincidence combinations among detector $d$ and all other detectors.
We performed a least squares fit of the detector efficiencies to match the average expected coverage and average observed coverage (average over the six data points \mbox{$C_\mathrm{AS}$, $C_\mathrm{BS}$, $C_\mathrm{CS}$, ... etc.}).
For the fitted procedure, the expected coverage $C_\ds$ was calculated from
detector efficiencies (Eq.~\ref{eq:coverage_d}) and exclusive cross-sections. 
We note that we evaluated exclusive cross-sections for detector-detector coincidences with the qubit array (\det{q}) omitted from the detector set.

We simplified the model by assuming all detectors share the same efficiency value, which is reasonable since $C_\ds$ does not depend strongly on any particular detector's efficiency.
This model simplification is also applicable for the coverage that the detectors provide the qubit array.
The fitted efficiency value shared among each detector is \mbox{$\epsilon_d=96\pm 3\%$}, where the uncertainty reflects the sample variance of the coverage.
We use the same sample variance of the coverage to estimate a fractional uncertainty of coverage \mbox{$\delta C/C=3\%$}.

We evaluated the muon flux \mbox{$\Phi=\valuePhi$} from the 
mean and sample variance of the six measured rates $r_\ds$ (Eq~\ref{eq:rate_any}).
In the qubit-detector coincidence experiment analysis, the muon flux in the laboratory was used to compare the cosmic-ray-induced \stc{} events rate with an expected rate that cosmic rays impact the qubit array.

\subsection{Application of the Rate Model to Qubit-detector Coincidences}\label{sec:app_rate_model}
\subsubsection{Calculation of coverage}\label{sec:coverage}
We used cross-sections and detector efficiencies to estimate the coverage that the detectors provide the qubit array.
We then calculated the rate of cosmic-ray-induced \stc{} qubit relaxation event from the coverage and the measured qubit-detector coincidence rate. 

The calculated cross-section for all cosmic-ray energy depositions to the qubit array is \mbox{$\sigmaCRQ=\valueSigmaQ{}$}.
A portion of cosmic rays also deposit energy in a detector:
The total cross-section for \det{qs} coincidences is the sum of exclusive cross-sections for all combinations that include the qubit array:
\begin{equation}
\sigmaCRQD = \sum_\alpha \sigma^*_\alpha = \valueSigmaQD{}
\end{equation}
where $\alpha$ sums over all combinations in the set
\mbox{\{\det{a}, \det{b}, \det{c}, \det{d}, \det{e}, \det{f}, \det{q}\}}
that include {\sc q} and a detector.
The expected occurrence rate of \det{qs} coincidences $r_\mathrm{QS}$ depends on the efficiency of the detectors within each \mbox{combination (Eq.~\ref{eq:rate_any}),}
\begin{equation}\label{eq:rateCRQD}
\rateCRQD=
\epCoin\epsilon_\mathrm{Q}
\overbrace{
\sum_\alpha
\bigg(1 - \prod_{
\substack{
k\in\alpha \\ k\neq\mathrm{Q}
}}
\overline{\epsilon}_k \bigg)
\sigma^*_\alpha
}^{\sigma_\mathrm{QS,\epsilon}}
\Phi
\end{equation}
where $k$ iterates over all detectors in the combination $\alpha$ (except \det{q}),  \mbox{$\overline{\epsilon}_k=1-\epsilon_k$} is the detector inefficiency, and \mbox{$\epsilon_\mathrm{Q}$} is an unknown cosmic-ray detection efficiency for the qubit array,
and we have also included $\epCoin=\valueEffQD$ as a known detection efficiency factor 
specific to coincidence identification within the coincidence window
(Section~\ref{sec:qb_det_coin}).

We refer to $\sigma_\mathrm{QS,\epsilon}$ as an effective cross-section that accounts for the collective efficiency of the detectors:
\mbox{$\epD = \sigma_\mathrm{QS,\epsilon} / \sigma_\mathrm{QS} = \valueEffS{}$}.
The coverage of the qubit array provided by the detectors is (Eq.~\ref{eq:coverage_d}),
\begin{equation}\label{eq:coverage_val}
\CQS=\frac{\rateCRQD}{\rateCRQ}
=\epCoin\epD\frac{\sigmaCRQD}{\sigmaCRQ} = \text{\valueCoverage{}},
\end{equation}
where the uncertainty is estimated from detector coverage measurements (Section~\ref{sec:eff_flux}).
The overall efficiency for these coincidences is \mbox{$\epQD=\epCoin\epD=\valueEff{}$}.
The coverage of other qubit-detector coincidence combinations (shown in Fig.~\ref{fig:fig3}) were calculated similarly, e.g.~\mbox{$C_\mathrm{QA}=r_\mathrm{QA}^\mu/\rateCRQ$}.

The coverage $\CQS$ was used to calculate the occurrence rate of \stc{} qubit relaxation events $\rateCRQ$ from cosmic rays:
\begin{equation}\label{eq:calc_rateCRQ}
\rateCRQ = \frac{1}{\CQS}\rateCRQD = \frac{1}{\CQS}(\rateQD - \bkgndQD)
=\frac{1}{\valueCRrateSeconds}
\end{equation}
where $\rateCRQD$ is the rate of qubit-detector coincidences from cosmic rays, $\rateQD$ is the rate of all observed qubit-detector coincidences, $\bkgndQD$ is the rate of accidental qubit-detector coincidences, and uncertainty was propagated from the uncertainty of the coverage (Eq.~\ref{eq:coverage_val}) and statistical uncertainty of $\rateQD$.

\subsection{Complete and Approximate Rate Calculation}\label{sec:rate_calc}
We applied our model of the cosmic ray flux to calculate the rate, or probability, for detector pulses and qubit events within an experiment observation window.
Here, we provide a framework to account for all possible processes that result in the observations predicted by our rate model.
While a single cosmic ray can cause multiple detectors to produce a pulse at the same time, there are other contributions to these observations that are not from individual cosmic rays, such as additional muon impacts that occur within the $\delta t$ observation window.
Here we provide a thorough evaluation of these contributions.

We consider the individual probabilities for each independent physical process that could result in observations of detector pulse combinations.
The probability of $n$ events from each independent process is given by the Poisson distribution,
$\pois{\lambda}{n} = \lambda^n e^{-\lambda} / n!$,
where $\lambda = r \delta t$ is the average number of events during the observation window $\delta t$.
For a given combination of detectors the occurrence rate $r$ is $\sigma^{*} \Phi$, where $\sigma^*$ is the exclusive cross-section and $\Phi$ is the muon flux (and we ignore detection efficiency for now).

The following decomposition gives an exact procedure to calculate the probability of observed detector pulse combinations within an observation window.
We provide this calculation method since it is a framework that can be extended for imperfect detection efficiencies, independent detector background pulses, and the addition of other particles fluxes.

First, we define a normalization expression which includes all possible and \emph{independent} physical processes.
Within an observation window $\delta t$, any number of rays could impact any of the detectors.
We decompose the normalization in terms of all exclusive combinations since
they each correspond to independent processes for cosmic rays traversing the detectors.
Since the Poisson distribution for each exclusive combination is normalized, we create the normalized expression
\begin{align}
1 & = \sum_{n=0}^{\infty} \pois{[\sigmaExA + \sigmaExB + \dots] \Phi \delta t}{n}                                                     \\
& = \sum_{n=0}^{\infty} \pois{\sigmaExA \Phi \delta t}{n} \times \sum_{m=0}^{\infty} \pois{\sigmaExB \Phi \delta t}{m} \times \dots\\
&=
\left(\sum_{n=0}^{n=\infty} \sum_{m=0}^{m=\infty} \dots \right)
\pois{\sigmaExA \Phi \delta t}{n} \times \pois{\sigmaExB \Phi \delta t}{m} \times \dots, \label{eq:unitarity}
\end{align}
where the product includes a sum for every exclusive cross-sections $\sigma_{\alpha^\prime}^*$.
Upon the expansion of terms in \mbox{Equation~\ref{eq:unitarity}}, each term has the physical meaning of \mbox{$n$, $m$, $\dots$} counts of cosmic rays impacting each of the detector combinations. 

We take a bottom-up approach to calculate the probability of an observed combination $\alpha$ within the observation window.
For a given experiment observation we apply selection functions that constrain the normalization expression and pick out all processes that would produce the observation.
Each constraint corresponds to a physical process that contributes to the total probability of an observation.
The observation probability for the combination $\alpha$ is
\begin{equation} \label{eq:unitarityWithConstraint}
P_\mathrm{\alpha} = 
\left(\sum_{n=0}^{n=\infty} \sum_{m=0}^{m=\infty} \dots \right)
\delta_{\alpha,\alpha^\prime}
\pois{\sigmaExA \Phi \delta t}{n} \times \pois{\sigmaExB \Phi \delta t}{m} \times \dots,
\end{equation}
where we denote $\alpha^\prime$ as the combination represented by each term upon an expansion of terms in \mbox{Equation~\ref{eq:unitarityWithConstraint}}.
Each combination $\alpha^\prime$ is formed by the union of the exclusive combinations within each these terms.

In practice, we list the independent processes that result in the observable $\alpha$ and then apply each constraint individually, adding the probability of each contribution.
We Taylor-expand \mbox{Equation~\ref{eq:unitarityWithConstraint}} based on orders of $\delta t$, finding the leading terms are the most likely physical processes to result in an observable $\alpha$.
For example, if there are only detectors \det{a} and \det{b} and we would like to know the probability for the exclusive coincidence \det{ab} to occur, we calculate
\begin{align*}
P_{\mathrm{AB}^{*}}
     & = 0 \times \pois{\lambda_{\mathrm{A}^{*}}}{0}\pois{\lambda_{\mathrm{B}^{*}}}{0}\pois{\lambda_{\mathrm{AB}^{*}}}{0}                                  \\
              & \qquad + 0 \times \pois{\lambda_{\mathrm{A}^{*}}}{1}\pois{\lambda_{\mathrm{B}^{*}}}{0}\pois{\lambda_{\mathrm{AB}^{*}}}{0}                           \\
              & \qquad + 0 \times \pois{\lambda_{\mathrm{A}^{*}}}{0}\pois{\lambda_{\mathrm{B}^{*}}}{1}\pois{\lambda_{\mathrm{AB}^{*}}}{0}                           \\
              & \qquad + 1 \times \pois{\lambda_{\mathrm{A}^{*}}}{0}\pois{\lambda_{\mathrm{B}^{*}}}{0}\pois{\lambda_{\mathrm{AB}^{*}}}{1} + \mathcal{O}(\delta t^2) \\
              & = \pois{\lambda_{\mathrm{A}^{*}}}{0}\pois{\lambda_{\mathrm{B}^{*}}}{0}\pois{\lambda_{\mathrm{AB}^{*}}}{1} + \mathcal{O}(\delta t^2)                 \\
              & \approx \lambda_{\mathrm{AB}^{*}}
\end{align*}
where $\lambda_{\alpha^{*}} = \sigma^{*}_\alpha \Phi \delta t$ are shorthand notations.
The terms that are second-order in $\delta t$ are less likely contributions to the observation $\det{ab}^{*}$,
such as \mbox{$\pois{\lambda_{\mathrm{A}^{*}}}{1}\pois{\lambda_{\mathrm{B}^{*}}}{1}\pois{\lambda_{\mathrm{AB}^{*}}}{0}$}
which corresponds to two separate cosmic rays impacting detectors \det{a} and \det{b}.

We calculate for imperfect detector efficiency by
separating each exclusive combination of the normalization expression \mbox{(Eq.~\ref{eq:unitarity})} into two independent processes that correspond to detection success and failure:
\mbox{$\lambda_{\alpha^{*}}\to(\epsilon + \overline{\epsilon})\lambda_{\alpha^{*}}$}, where \mbox{$\overline{\epsilon}=1-\epsilon$}.
We generalize the probability to observe a combination $\alpha$ (Eq.~\ref{eq:unitarityWithConstraint}) as,
\begin{equation} \label{eq:unitarityWithConstraintEff}
P_\mathrm{\alpha}=\left(\sum_{n=0}^{n=\infty} \sum_{m=0}^{m=\infty} \dots \right)
\delta_{\alpha,\alpha^\prime}
\mathrm{Pois}\big((\epsilon_\mathrm{A}+\overline{\epsilon}_\mathrm{A})\sigmaExA \Phi \delta t\big)_{n} \times
\mathrm{Pois}\big((\epsilon_\mathrm{B}+\overline{\epsilon}_\mathrm{B})\sigmaExB \Phi \delta t\big)_{m} \times \dots.
\end{equation}
The expanded expression of Eq.~\ref{eq:unitarityWithConstraintEff} now has terms for
both when detectors have a pulse (\mbox{$\sim\epsilon$}) and when detectors fail to pulse (\mbox{$\sim\overline{\epsilon}$}).
The constrains applied by each $\delta_{\alpha,\alpha^\prime}$ now also account for 
the requirement for specific detectors to pulse (or fail to pulse) in order to retain terms that match the observation $\alpha$.
For example, the probability of observing a coincidence \det{ab} exclusively within $\delta t$ (again, in the case of the detector set \det{a} and \det{b}):
\begin{align*}
P_{\mathrm{AB}^{*}}
& = 0 \times \pois{\lambda_{\mathrm{A}^{*}}}{0}\pois{\lambda_{\mathrm{B}^{*}}}{0}\pois{\lambda_{\mathrm{AB}^{*}}}{0}\\
& \qquad + \left[0 \times \epsilon_\mathrm{A} + 0 \times \overline{\epsilon}_\mathrm{A} \right] \pois{\lambda_{\mathrm{A}^{*}}}{1}\pois{\lambda_{\mathrm{B}^{*}}}{0}\pois{\lambda_{\mathrm{AB}^{*}}}{0}\\
& \qquad + \left[0 \times \epsilon_\mathrm{B} + 0 \times \overline{\epsilon}_\mathrm{B} \right] \pois{\lambda_{\mathrm{A}^{*}}}{0}\pois{\lambda_{\mathrm{B}^{*}}}{1}\pois{\lambda_{\mathrm{AB}^{*}}}{0}\\
& \qquad + \left[1 \times \epsilon_\mathrm{A} \epsilon_\mathrm{B} + 0 \times \epsilon_\mathrm{A} \overline{\epsilon}_\mathrm{B} + 0 \times \overline{\epsilon}_\mathrm{A} \epsilon_\mathrm{B} + 0 \times \overline{\epsilon}_\mathrm{A} \overline{\epsilon}_\mathrm{B} \right] \pois{\lambda_{\mathrm{A}^{*}}}{0}\pois{\lambda_{\mathrm{B}^{*}}}{0}\pois{\lambda_{\mathrm{AB}^{*}}}{1} \\
& \qquad + \mathcal{O}(\delta t^2)\\
& = \epsilon_\mathrm{A} \epsilon_\mathrm{B} \pois{\lambda_{\mathrm{A}^{*}}}{0}\pois{\lambda_{\mathrm{B}^{*}}}{0}\pois{\lambda_{\mathrm{AB}^{*}}}{1} + \mathcal{O}(\delta t^2)\\
& \approx \epsilon_\mathrm{A} \epsilon_\mathrm{B} \lambda_{\mathrm{AB}^{*}}
\end{align*}
Since $\lambda \lesssim 10^{-3}$ for all occurrence rates and observation windows used throughout our experiment analysis, we generally approximated occurrence rates based on first-order (in $\delta t$) contributions to the probability of observing an event in each observation window~\mbox{(Eq.~\ref{eq:rate_approx2})}.
The second-order terms correspond to an additional one cosmic-ray energy deposition for every $10^6$ observed, which is an insignificant correction compared to other possible discrepancies between this rate model and observations of cosmogenic particles.

\end{document}